\documentclass[final,3p,times,authoryear]{elsarticle}
\usepackage{amssymb}

\usepackage{amsmath,amsthm,mathtools}
\usepackage{amsfonts}
\usepackage{bm}

\usepackage{bbm}
\DeclareMathOperator*{\esssup}{ess\,sup}
\usepackage{caption}
\usepackage{subcaption}

\newtheorem{thm}{Theorem}

\usepackage{verbatim}

\usepackage{hyperref}
\hypersetup{
    colorlinks=true,
    linkcolor=blue,
    citecolor=blue,
    filecolor=magenta,   
    urlcolor=blue, %cyan
    pdftitle={Overleaf Example},
    pdfpagemode=FullScreen,
}

\usepackage{booktabs}

\begin{document}

%%%%%%%%%%%%%%%%%%%%%%%%%%%%%%%%%%%%%%%%%%%%%%%%%%%%%%%%%%%%%%

\begin{frontmatter}

\title{Conformal Prediction Bands for Two-Dimensional Functional Time Series}

\author[MOX]{Niccolò Ajroldi}
\author[Padova]{Jacopo Diquigiovanni}
\author[JRC]{Matteo Fontana\corref{cor1}}
\author[MOX]{Simone Vantini}

\cortext[cor1]{Email: matteo.fontana@ec.europa.eu}

\affiliation[MOX]{organization={MOX - Department of Mathematics, Politecnico di Milano},
            city={Milan (MI)},
            country={Italy}}
\affiliation[Padova]{organization={Department of Statistical Sciences, University of Padova},
            city={Padova (PD)},
            country={Italy}}
\affiliation[JRC]{organization={European Commission, Joint Research Centre (JRC)},
            city={Ispra (VA)},
            country={Italy}}

\begin{abstract}
    Time evolving surfaces can be modeled as two-dimensional Functional time series, exploiting the tools of Functional data analysis.
    Leveraging this approach, a forecasting framework for such complex data is developed.
    The main focus revolves around Conformal Prediction, a versatile nonparametric paradigm used to quantify uncertainty in prediction problems.
    %This work focuses in particular on Conformal Prediction, a versatile nonparametric framework used to quantify uncertainty in prediction problems.
    Building upon recent variations of Conformal Prediction for Functional time series, a probabilistic forecasting scheme for two-dimensional functional time series is presented, while providing an extension of Functional Autoregressive Processes of order one to this setting.
    Estimation techniques for the latter process are introduced, and their performance are compared in terms of the resulting prediction regions.
    Finally, the proposed forecasting procedure and the uncertainty quantification technique are applied to a real dataset, collecting daily observations of Sea Level Anomalies of the Black Sea.
\end{abstract}

% \begin{abstract}
%     Conformal Prediction (CP) is a versatile nonparametric framework used to quantify uncertainty in prediction problems.
%     In this work, we provide an extension of such method to time series of functions defined on a bivariate domain, 
%     proposing a distribution-free technique which can be applied to time-evolving surfaces.
%     In order to obtain meaningful and efficient prediction regions, CP must be coupled with an accurate forecasting algorithm, for this reason, we extend the theory of autoregressive processes in Hilbert space in order to allow for functions with a bivariate domain.
%     Given the novelty of the subject, we present estimation techniques for the Functional Autoregressive model (FAR). % and for principal component analysis (PCA) for %two-dimensional functional data. 
%     A simulation study is implemented, in order to investigate how different point predictors affect the resulting prediction bands.
%     Finally, we explore benefits and limits of the proposed approach on a real dataset, collecting daily observations of Sea Level Anomalies of the Black Sea in the last twenty years. \\
% \end{abstract}


\end{frontmatter}

%%%%%%%%%%%%%%%%%%%%%%%%%%%%%%%%%%%%%%%%%%%%%%%%%%%%%%%%%%%%%%

%{
%  \hypersetup{linkcolor=black}
%  \tableofcontents
%}
%\newpage

%%%%%%%%%%%%%%%%%%%%%%%%%%%%%%%%%%%%%%%%%%%%%%%%%%%%%%%%%%%%%%

% \newpage
\section{Introduction}
\label{section:intro}
    
    Data observed on a two-dimensional domain arise naturally across several disciplines, motivating an increasing demand for dedicated analysis techniques.
    Functional data analysis (FDA) (\citealt{ramsay2005functional}) is naturally apt to represent and model this kind of data, as it allows preserving their continuous nature, and provides a rigorous mathematical framework. 
    Among the others, 
    \citealt{Zhou} analyzed temperature surfaces, presenting two approaches for Functional Principal Component Analysis (FPCA) of functions defined on a non-rectangular domain,
    \citealt{Munoz} focuses on image processing using FDA, %proposing a representation of iris images through functional data, 
    while a novel regularization technique for Gaussian random fields on a rectangular domain has been proposed by \citealt{Raket} and applied to 2D electrophoresis images. 
    Another bivariate smoothing approach in a penalized regression framework 
    has been introduced by \citealt{Ivanescu}, allowing for the estimation of functional parameters of two-dimensional functional data.
    As shown by \citealt{Gervini}, even mortality rates can be interpreted as two-dimensional functional data.%, where one dimension is the temporal one ad the other one refers to age.
    
    Whereas in all the reviewed works functions are assumed to be realization of \textit{iid} or at least \textit{exchangeable} random objects, to the best of our knowledge there is no literature focusing on forecasting time-dependent two-dimensional functional data.
    In this work, we focus on time series of surfaces, representing them as two-dimensional Functional Time Series (FTS).
    
    A two-dimensional Functional Time Series is an ordered sequence $Y_1, \dots, Y_T$ of random variables with values in a functional Hilbert space $\mathbb{H}$, characterized by some sort of temporal dependency.
    More formally, we consider a probability space $(\Omega, \mathcal{F},\mathbb{P})$, and define a random function at time $t$ as $Y_t: \Omega \rightarrow \mathbb{H}$, measurable with respect to the Borel $\sigma$-algebra $\mathcal{B}(\mathbb{H})$. 
    In the rest of the article we consider functions belonging to $\mathbb{H} = \mathcal{L}^2([c,d] \times [e,f])$, 
    %the space of measurable square integrable real-valued functions defined on the rectangle $[c,d] \times [e,f] \subset \mathbb{R}^2$, 
    with $c,d,e,f \in \mathbb{R}$, $c<d$, $e<f$
    %\footnote{Such choice is motivated by many reasons, one above all is the fact that, by considering functions in $\mathcal{L}^2([c,d] \times [e,f])$, the usual Fréchet mean for functional data coincides with the pointwise mean and the covariance kernel coincides with the point-wise covariance. 
    %Moreover, $\mathbb{H}$ is a separable Hilbert space, with the usual inner product:
    %$
       %\langle x,y \rangle := \int_c^d \int_e^f x(u,v) y(u,v) du dv \qquad \forall x,y \in \mathbb{H} 
    %$
    %}.
    We stress the fact that, from a theoretical point of view, our methodology can be applied to functions defined on a generic subset of $\mathbb{R}^2$, however, for simplicity and without loss of generality, we will only consider rectangular domains.
    
    Given a realization of the stochastic process $\{Y_t\}_{t= \, \dots, N}$, we aim to forecast the next surface and quantify the uncertainty around the predicted function.
    Whereas uncertainty quantification in the context of \textit{univariate} FTS forecasting has received great attention in the statistical community in recent decades, no attempts have been made to extend them to functions defined on a bidimensional domain.
    %Among the different publications tackling the problem of distributional forecasting, some authors like \citealt{Hyndman_Ullah} relies on parametric assumptions, while many others have developed  extension of the Bootstrap (see e.g. \citealt{Hyndman_Shang}, \citealt{Rossini} and more recently also \citealt{hernandez2021simultaneous}).
    %Among the different publications tackling the problem of quantifying uncertainty in FTS forecasting, 
    
    Most of the research tackling univariate FTS forecasting has focused on adaption of the Bootstrap to the functional setting (see e.g. \citealt{Hyndman_Shang}, \citealt{Rossini} and \citealt{hernandez2021simultaneous}). However, Bootstrap is a very computationally intensive procedure, especially in the infinite-dimensional context of functional data.
    In this work, we instead focus on Conformal Prediction (CP), a versatile nonparametric approach to prediction. The first appearance of such technique dates back to \citealt{gammerman}, and it has been later presented in greater detail in \citealt{Vovk2005AlgorithmicLI} and \citealt{Balasubramanian}. 
    An extensive and unified review of the theory of Conformal inference can be found in \citealt{Zeni2020ConformalPA}.
    The attractiveness of Conformal Prediction relies on its great versatility, which allows to couple it with any predictive algorithm, in order to obtain distribution free prediction sets.     
    Throughout this work, we resort to Inductive Conformal Prediction, also known as Split Conformal Prediction (\citealt{Papadopoulos2002InductiveCM}). Such modification of the original Transductive Conformal method is not only computationally efficient, but also necessary in high-dimensional frameworks like the functional data one.
    It should be noted that the main drawback of the Full Conformal approach is the need of retraining the prediction algorithm for every possible candidate realization $y$. In practice, in multivariate problems, where $y$ lies in $\mathbb{R}^p$, one runs the above routine for several candidates $y$ over a $p$-dimensional regular grid.
    While such approach is prohibitive for high-dimensional spaces, since computational times grow exponentially with $p$, it becomes unfeasible in a functional setting, in which $y$ lies in an infinite-dimensional space.
    Employing Split Conformal inference along with nonconformity scores tailored to functional data (\citealt{diquigiovanni2021importance}) allows to obtain prediction sets in closed form.
    
    Whereas CP theory has been originally developed under the assumption of exchangeability, \citealt{chernozhukov2018exact} reframed the CP framework in the context of randomization inference, proving approximate validity of the resulting prediction sets under weak assumptions on the conformity scores and on the ergodicity of the time series. Later, \citealt{diquigiovanni2021_FTS} adapted such methodology to allow for Functional Time Series in a Split Conformal setting.
    We extend such method to two-dimensional functional data. 
    
    Since we want to quantify prediction uncertainty, we also need to provide a \textit{forecasting technique} for 2D Functional Time Series.
    We start by taking into account the literature on unidimensional FTS forecasting (\citealt{bosq}, \citealt{antoniadis}, \citealt{horvath2012inference}, \citealt{Aue}, \citealt{jiao}), extending the theory of Functional Autoregressive Processes (FAR) to the bivariate setting, proposing estimation techniques for the FAR(1), and comparing them in an order to assess how forecasting performances influence the amplitude of prediction bands.
    
    The rest of this paper is as follows: we illustrate Conformal Prediction for two-dimensional functional data in Section \ref{section:CP}, providing theoretical guarantees of the resulting prediction bands.  
    In Section \ref{section:point_prediction}, we introduce forecasting algorithms for two-dimensional Functional Time Series, proposing an extension of the FAR(1) for two-dimensional functional data.
    Such forecasting algorithms are then 
    %adapted to the Conformal inference setting and consequently 
    compared by means of the resulting prediction bands in a simulation study in Section \ref{section:4}. 
    Finally, in Section \ref{section:BlackSea} we employ the developed techniques to obtain forecasts and prediction bands of real data, predicting day by day the Black Sea level. Section \ref{section:conclusions} concludes.
    
%%%%%%%%%%%%%%%%%%%%%%%%%%%%%%%%%%%%%%%%%%%%%%%%%%%%%%%%%%%%%%
%%%%%%%%%%%%%%%%%%%%%%%%%%%%%%%%%%%%%%%%%%%%%%%%%%%%%%%%%%%%%%

%\section{Conformal Inference for Functional Time \linebreak Series}
%\section{Conformal Prediction Bands for Functional \linebreak Time Series}

%%%%%%%%%%%%%%%%%%%%%%%%%%%%%%%%%%%%%%%%%%%%%%%%%%%%%%%%%%%%%%

%\section{Uncertainty quantification methodology: Conformal Prediction}
%\section{Methodology: Conformal Prediction for 2D Functional Data}
% \section{Methodology: Conformal Prediction for 2D Functional Time Series}
\section{Uncertainty Quantification: Conformal Prediction for 2D Functional Time Series}

%s\section{Uncertainty quantification methodology}
\label{section:CP}
    
    Consider a time series $Z_1, \dots, Z_T$ of regression pairs $Z_t = (X_t, Y_t)$, with $t=1, \dots, T$. 
    Let $Y_t$ be a random variable with values in $\mathbb{H}$, while $X_t$ is a set of random covariates at time $t$ belonging to a measurable space. Notice that $X_t$ is a generic set of regressors, which may contain both exogenous and endogenous variables. Later in the manuscript, we will consider $X_t$ to contain only the lagged version of the function $Y_t$, namely $Y_{t-1}$.
    Given a significance level $\alpha \in [0,1]$, we aim to design a procedure that outputs a prediction set $\mathcal{C}_{T,1-\alpha}(X_{T+1})$ for $Y_{T+1}$ based on $Z_1, \dots, Z_{T}$ and $X_{T+1}$,
    with unconditional coverage probability close to $1-\alpha$.
    More formally, we define $\mathcal{C}_{T,1-\alpha}(X_{T+1})$ to be a \textit{valid} prediction set if:
    \begin{equation}
    \label{eqn:valid_set}
        \mathbb{P}(Y_{T+1} \in \mathcal{C}_{T,1-\alpha}(X_{T+1})) \geq 1 - \alpha.
    \end{equation}
    % for any probability distribution $P$ and ...
    We would like to construct a specific type of prediction sets, commonly known as \textit{prediction bands}, formally defined as:
    \begin{equation}
        \label{eq:bands}
        \{ y \in \mathbb{H}: y(u,v) \in B_{n}(u,v) \quad \forall (u,v) \in [c,d] \times [e,f] \},
    \end{equation}
    where $B_n(u,v) \subseteq \mathbb{R}$ is an interval for each $(u,v) \in [c,d] \times [e,f]$. 
    The convenience of such type of prediction sets in applications is extensively motivated in the literature (see e.g. \citealt{Pintado}, \citealt{Lei1} and \citealt{diquigiovanni2021importance}), since a prediction set of this kind can be visualized easily, a property that is instead not guaranteed if the prediction region is a generic subset of $\mathbb{H}$.
    
    Let $z_1, \dots, z_T$ be realizations of $Z_1, \dots, Z_T$. 
    As the name suggests, Split Conformal inference is based on a random split of data into two disjoint sets: let $\mathcal{I}_1$, $\mathcal{I}_2$ be a random partition of $\{1, \dots, T\}$, such that $|\mathcal{I}_1|=m$, $|\mathcal{I}_2|=l$, $m,l \in \mathbb{N}$ $m,l > 0$, $m+l=T$. 
    Historical observations $z_1, \dots, z_T$ are divided into a \textit{training set} $\{z_h,\, h \in \mathcal{I}_1\}$, used for model estimation, and a \textit{calibration set} $\{z_h,\, h \in \mathcal{I}_2\}$, used in an out-of-sample context to measure the nonconformity of a new candidate function. 
    The choice of the split ratio and the type of split is non-trivial and has motivated discussion in the statistical community. We fix the training-calibration ratio equal to 1 and perform a random split, and refer to \ref{appendix_split} for a more extensive discussion on the topic.
    
    We then introduce a \textit{nonconformity measure} $\mathcal{A}$, which is a measurable function with values in $\mathbb{R} \cup \{+\infty\}$. 
    The role of $\mathcal{A}(\{z_h, \, h \in \mathcal{I}_1\}, z)$ is to quantify the nonconformity of a new datum $z$ with respect to the training set $\{z_h, \, h \in \mathcal{I}_1\}$. 
    The choice of the nonconformity measure is crucial to find prediction bands \eqref{eq:bands} in closed form. 
    % Moreover, as will be discussed later, this choice also affects the size of the resulting prediction bands, and thus their efficiency.
    % Motivated by such considerations, we employ the following nonconformity score, introduced by \citealt{diquigiovanni2021importance}, extended here to two-dimensional functional data:
    We employ the following nonconformity score, introduced by \citealt{diquigiovanni2021importance}, extended here to two-dimensional functional data:
    \begin{equation}
    \label{eqn:nonconformitymeasure}
        \mathcal{A}(\{z_h: h \in \mathcal{I}_1\}, z) = \esssup_{(u,v) \in [c,d] \times [e,f]} \frac{| y(u,v) - g_{\mathcal{I}_1}(u,v;x_{T+1}) |}{s_{\mathcal{I}_1}(u,v)},
    \end{equation}
    where $z=(x_{T+1}, y)$, $g_{\mathcal{I}_1}$ is a point predictor built from the training set $\mathcal{I}_1$,  and $s_{\mathcal{I}_1}$ is a \textit{modulation function}, which is a positive function depending on $\mathcal{I}_1$ that allows for prediction bands with non-constant width along the domain. 
    Section \ref{section:4} and \ref{appendix_EK} discuss the estimation of $g_{\mathcal{I}_1}$.
    % The estimation of $g_{\mathcal{I}_1}$ is discussed in Section \ref{section:4}, 
    The functional standard deviation is employed as modulation function $s_{\mathcal{I}_1}$, allowing for wider bands in the parts of the domain where data show high variability and narrower and more informative prediction bands in those parts characterized by low variability. For an extensive discussion on the optimal choice of modulation function, we refer to \citealt{diquigiovanni2021importance}.
    \begin{comment}
    Notice that, from a theoretical point of view, the nonconformity measure $\mathcal{A}$ may be $+\infty$,
    since we are embedding functions in $\mathcal{L}^2([c,d] \times [e,f])$ and we have thus no guarantee on their boundedness. To overcome this issue, one can instead consider the functional space $\mathcal{L}^{\infty}([c,d] \times [e,f])$, as done by \citealt{diquigiovanni2021importance}, however, such space equipped with the usual $\mathcal{L}^2$ scalar product, is not closed, and is therefore not a Hilbert space. 
    For such reason, we decided to settle anyway the analysis in $\mathcal{L}^2([c,d] \times [e,f])$, resorting to the fact that, in practical applications, \eqref{eqn:nonconformitymeasure} will only assume finite values, given the finite nature of observed data.
    \end{comment}
    
    Consider now a candidate function $y \in \mathbb{H}$ and define the augmented dataset as $Z_{(y)} = \{Z_t\}_{t=1}^{T+1}$, where:
    \begin{equation}
        Z_t = \begin{cases}
             (X_t, Y_t),& \text{if } 1 \leq t \leq T, \\
             (X_{T+1}, y) ,& \text{if } t = T+1.
        \end{cases}
    \end{equation}
    The key idea of the methodology proposed by \citealt{chernozhukov2018exact} and extended by \citealt{diquigiovanni2021_FTS} is to generate several randomized versions of $Z_{(y)}$ through a specifically tailored permutation scheme, and compute nonconformity scores on each of them.
    We then decide whether to include $y$ in the prediction region, by comparing the nonconformity score of $Z_{(y)}$ with that of its permuted replicas.
    %randomized versions.
    
    In order to obtain such replicas, we aim to define a family $\Pi$ of index permutations $\pi: \{1,\dots,T+1\} \rightarrow \{1,\dots,T+1\}$, that keeps unchanged the training set indices $\mathcal{I}_1$, and modifies only $\{\mathcal{I}_2, T+1\}$, namely the indices of the calibration set and the next time step.
    % In order to do so, let's 
    We first introduce a function $\lambda: \{\mathcal{I}_2, T+1\} \rightarrow \{1,\dots,l+1\}$ such that $\lambda(t)$ returns the $t$-th element of the ordered set $\{\mathcal{I}_2,T+1\}$. 
    Fix now a positive integer $b \in \{1, \dots l+1\}$ such that $\frac{l+1}{b} \in \mathbb{N}$ and define a family $\tilde{\Pi}$ of index permutations that acts on the set $\{1,\dots,l+1\}$. 
    Each $\tilde{\pi}_i \in \tilde{\Pi}$ is required to be a bijection $\tilde{\pi}_i : \{1,\dots,l+1\} \rightarrow \{1,\dots,l+1\}$, for $i = 1, \dots, \frac{l+1}{b}$. 
    %In particular, we will consider non-overlapping blocking permutations (\citealt{chernozhukov2018exact}), dividing data in blocks of size $b$ (notice that each permutation is unique in the group $\tilde{\Pi}$, from which the name "):
    We consider the non-overlapping blocking permutation scheme proposed by \citealt{chernozhukov2018exact}, dividing data in blocks of size $b$, in such a way that each permutation is unique in $\tilde{\Pi}$:
    \begin{equation}
        \tilde{\pi}_i(j) = 
        \begin{cases}
             j+(i-1)b & \text{if } 1 \leq j \leq l - (i-1)b+1, \\
             j+(i-1)b-l-1 & \text{if } l - (i-1)b+2 \leq j \leq l+1.
         \end{cases}
    \end{equation}
    By definition, we have that $|\tilde{\Pi}|= \frac{l+1}{b}$, and $\tilde{\Pi}$ forms an algebraic group, containing the identity transformation $\tilde{\pi}_1$.
    It is then straightforward to introduce the family $\Pi$ of index permutations acting on $\{1,\dots,T+1\}$. Each $\pi_i \in \Pi$, with $i=1,\dots,\frac{l+1}{b}$ is defined as:
    \begin{equation}
    \label{eqn:blocking_scheme}
        \pi_i(t) = 
        \begin{cases}
             t & \text{if } t \in \mathcal{I}_1, \\
             \lambda^{-1}(\tilde{\pi}_i(\lambda(t)))) & \text{if } t \in \mathcal{I}_2 \cup \{T+1\}.
         \end{cases}
    \end{equation}
    
    \begin{figure}
        \centering \includegraphics[width=0.65\linewidth]{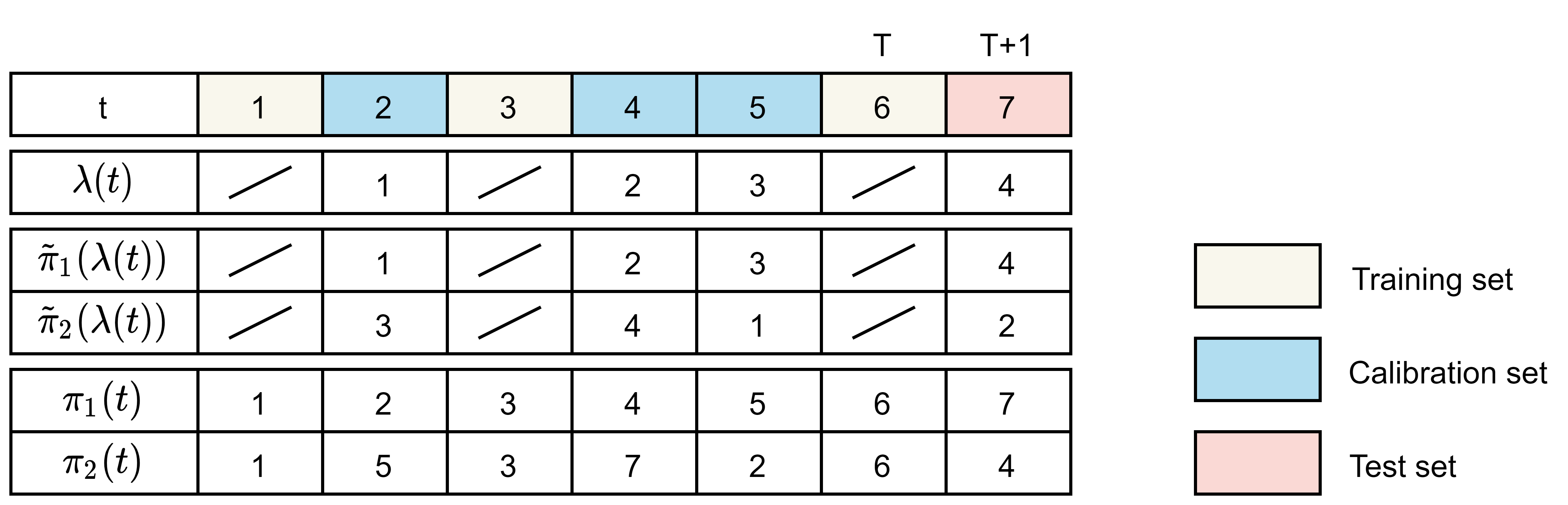}
        \caption[Example of permutation families $\tilde{\Pi}$ and $\Pi$]{Example of permutation families $\tilde{\Pi}$ and $\Pi$, with sample size $T=6$, training set $\mathcal{I}_1=\{1,3,6\}$, calibration set $\mathcal{I}_2=\{2,4,5\}$, $l=m=3$, size of blocking scheme $b=2$. In this case $\lambda:\{\mathcal{I}_2,T+1\}\equiv\{2,4,5,7\}\rightarrow \{1,2,3,4\}$, $\tilde{\Pi}=\{\tilde{\pi}_1,\tilde{\pi}_2\}$ and $\Pi=\{\pi_1,\pi_2\}$.}
        \label{fig:permutations}
    \end{figure}
    Figure \ref{fig:permutations} reports an example of the families of permutation $\Pi$ and $\tilde{\Pi}$.
    We refer to $Z^{\pi}_{(y)} = \{Z_{\pi(t)}\}_{t=1}^{T+1}$ as the randomized version of $Z_{(y)}=\{Z_t\}_{t=1}^{T+1}$, and define the randomization p-value as:
    \begin{equation}
    \label{eqn:p_value_function}
        p(y) = \frac{1}{|\Pi|} \sum_{\pi \in \Pi} \bm{1} (S(Z^{\pi}_{(y)}) \geq S(Z_{(y)})),
    \end{equation}
    where nonconformity scores $S(Z_{(y)})$ and $S(Z^{\pi}_{(y)})$ are defined as:
    \begin{gather}
        S(Z_{(y)}) = \mathcal{A}(\{Z_h\!: h \in \mathcal{I}_1\}, Z_{T+1}), \\
        S(Z^{\pi}_{(y)}) = \mathcal{A}(\{Z_h\!: h \in \mathcal{I}_1\}, Z_{\pi(T+1)}).
    \end{gather}
    %\textcolor{red}{Da capire se qui uso Z maiuscolo (rv) o minuscolo (realization). Se uso $\hat{p}$ e non $p$ allora forse devo usare le minuscole.} \\
    
    The idea is to apply permutations, modifying the order of observations in the calibration set, while at the same time preserving the dependence between them, thanks to the block structure of $\Pi$. For each $\pi$, we compute the nonconformity score of $Z^{\pi}_{(y)}$. The p-value \ref{eqn:p_value_function} of a test candidate value $y$ is then determined %by comparing the nonconformity score of the candidate $y$ with the scores of the $y$ after having applied the transformation $\pi$.
    as the proportion of randomized versions $Z^{\pi}_{(y)}$ with a higher or equal nonconformity score than the one of the original augmented dataset $Z_{(y)}$.
    Notice that $p(y)$ is a measure of the \textit{conformity} of the candidate function $y$ with respect to the permutation family $\Pi$. 
    It is then natural to include in the prediction set only functions $y$ with an ``high" conformity level.
    %Therefore, in accordance with Conformal Prediction, given a significance level $\alpha \in [b/(l+1),1]$\footnote{If $\alpha \in (0,b/(l+1))$ the resulting prediction set coincide with the entire space $\mathbb{H}$ (\citealt{diquigiovanni2021importance}).}, we define the prediction bands by test inversion:
    Given a significance level $\alpha \in [b/(l+1),1]$, the prediction region is hence obtained by test inversion:
    \begin{equation}
        \mathcal{C}_{T,1-\alpha}(X_{T+1}) := \{y \in \mathbb{H}\!: \, p(y) > \alpha\}.
    \end{equation}
As a remark, it should be noted that if $\alpha \in (0,b/(l+1))$ the resulting prediction set coincides with the entire space $\mathbb{H}$ (\citealt{diquigiovanni2021importance}).
    
    The advantage of using the Split Conformal method along with the conformity measure \eqref{eqn:nonconformitymeasure} relies on the possibility to find the prediction set in closed form.
    By defining $k^s$ as the $\lceil (|\Pi|+1)(1-\alpha) \rceil$th smallest value of $\{S(Z_{(y)}^{\pi}), \pi \in \Pi \setminus \pi_1\}$, we derive:% a prediction band in closed form:
    
    \begin{gather*}
        y \in \mathcal{C}_{T,1-\alpha}(X_{T+1}) \iff p(y) > \alpha
        \iff S(Z_{(y)}) \leq k^s \\
        \iff \esssup_{(u,v) \in [c,d] \times [e,f]} \frac{| y(u,v) - g_{\mathcal{I}_1}(u,v;x_{T+1}) |}{s_{\mathcal{I}_1}(u,v)} \leq k^s \\
        \iff | y(u,v) - g_{\mathcal{I}_1}(u,v;x_{T+1}) | \leq k^s s_{\mathcal{I}_1}(u,v) \qquad \forall (u,v) \in [c,d] \times [e,f] \\
        \iff y(u,v) \in \left[ g_{\mathcal{I}_1}(u,v;x_{T+1}) \pm k^s s_{\mathcal{I}_1}(u,v) \right] \qquad \forall (u,v) \in [c,d] \times [e,f].
    \end{gather*}
    %Therefore, the prediction set in closed form is defined as:
    The prediction band is therefore:
    \begin{equation}
    \label{eqn:prediction_set_explicit}    
        \mathcal{C}_{T,1-\alpha}(X_{T+1}):= \left\{y \in \mathbb{H}: y(u,v) \in \left[ g_{\mathcal{I}_1}(u,v;x_{T+1}) \pm k^s s_{\mathcal{I}_1}(u,v) \right] \forall (u,v) \in [c,d] \times [e,f] \right\}.
    \end{equation}
    
    If regression pairs are exchangeable, the proposed method retains exact, model-free validity (\citealt{chernozhukov2018exact}, Theorem 1). 
    When such assumption is not met, one can instead guarantee approximate validity of the proposed approach under weak assumptions on the nonconformity score and the ergodicity of the time series. 
    This result is illustrated in great detail by Theorem 2 of \citealt{chernozhukov2018exact} and Theorem 1 of \citealt{diquigiovanni2021_FTS}. We report here the latter, with a slightly modified notation. 
    
    Let $Z = Z_{(Y_{T+1})}$, where the candidate function $y$ is now substituted by the random function $Y_{T+1}$.
    Let $\mathcal{A}^*$ be an oracle nonconformity measure, inducing oracle nonconformity score $S^*$. Define $F$ to be the cumulative (unconditional) distribution function of the oracle nonconformity scores, namely $F(x)=\mathbb{P}(S^*(Z_{(y)}^{\pi})<x)$ and $\hat{F}$ the empirical counterpart, obtained by applying permutations $\pi \in \Pi$: $\hat{F}(x) := \frac{1}{|\Pi|} \sum_{\pi \in \Pi} \mathbbm{1}\{S^*(Z_{(y)}^{\pi}) < x\}$. 
    Let $\{\delta_{1\bar{l}},\delta_{2\bar{m}},\gamma_{1\bar{l}},\gamma_{2\bar{m}}\}$ be sequences of numbers converging to zero.

    \begin{thm}
    \label{my_theorem}
        If the following conditions hold:
        \begin{itemize}
            \item $\sup_{a \in \mathbb{R}} |\hat{F}(a)-F(a)| \leq \delta_{1\bar{l}}$ with probability $1-\gamma_{1\bar{l}}$
            \item $\frac{1}{|\Pi|} \sum_{\pi \in \Pi}\left[S(Z^{\pi})-S^*(Z_{(y)}^{\pi})\right]^2 \leq \delta_{2\bar{m}}^2$ with probability $1-\gamma_{2\bar{m}}$
            \item $|S(Z^{\pi}) - S^*(Z^{\pi})| \leq \delta_{2\bar{m}}$ with probability $1-\gamma_{2\bar{m}}$
            \item  With probability $1-\gamma_{2\bar{m}}$ the pdf of $S^*(Z^{\pi})$ is bounded above by a constant $D$
        \end{itemize}
        then the Conformal confidence set has approximate coverage $\-\alpha$:
        \begin{equation}
            |\mathbb{P}(Y_{T+1}\in \mathcal{C}_{T,1-\alpha}(X_{T+1}))-(1-\alpha)| \leq 6 \delta_{1\bar{l}}+ 2 \delta_{2\bar{m}} + 2 D \left( \delta_{2\bar{m}} + 2 \sqrt{\delta_{2\bar{m}}} \right) + \gamma_{1\bar{l}} + \gamma_{2\bar{m}}.
        \end{equation}
    \end{thm}
    
    The first condition concerns the approximate ergodicity of $\hat{F}$, a condition which holds for strongly mixing time series using blocking permutation $\Pi$ defined in \eqref{eqn:blocking_scheme} (\citealt{chernozhukov2018exact}).
    The other conditions are requirements for the quality of approximation of $S^*(Z^{\pi})$ with $S(Z^{\pi})$. Intuitively, $\delta_{2\bar{m}}^2$ bounds the discrepancy between the nonconformity scores and their oracle counterparts. Such condition is related to the quality of the point prediction and to the choice of the employed nonconformity measure.
    
%%%%%%%%%%%%%%%%%%%%%%%%%%%%%%%%%%%%%%%%%%%%%%%%%%%%%%%%%%%%%%

% \section{Point prediction}
\section{Point Prediction: Functional Autoregressive Process of order one}
\label{section:point_prediction}
    
    In order to obtain CP band with empirical coverage close to the nominal one, %while maintaining a small width,
    the choice of an accurate point predictor is important. As mentioned before, whereas in the typical i.i.d. case finite-sample unconditional coverage still holds when the model is heavily misspecified (\citealt{diquigiovanni2021importance}), in the time series context a strong model misspecification may compromise the coverage guarantees and not only the efficiency of the resulting prediction bands (\citealt{chernozhukov2018exact}, \citealt{diquigiovanni2021_FTS}). 
    For this reason, it is important to consider models that are consistent with the functional nature of the observations and that can adequately deal with their infinite dimensionality.
    We build on top of the literature on functional autoregressive processes in Hilbert spaces, extending them for the first time to temporarily evolving surfaces.
    We  narrow the forecasting methodology to the FAR(p), with $p=1$ because of its wide success in the literature (\citealt{hernandez2021simultaneous}, \citealt{Papadopoulos2002InductiveCM} and \citealt{Aue}).
    Whereas in the scalar context it is often beneficial to consider lags $p$ greater than one, given the intrinsic high dimensionality of functional data, we would rather fit a biased but simpler model than an unbiased but more complicated model.
    This issue is enhanced in the two-dimensional context, because of the extra dimension in the domain of the function, and for this reason, we consider only the case $p=1$.
    We introduce the Functional Autoregressive model of order 1 in Section \ref{sec:FAR1} and propose estimation techniques in Section \ref{sec:FAR1_estimation}.
    %%%%%%%%%%%%%%%%%%%%%%%%%%%%%%%%%%%%%%%%%%%%%%%%%%%%%%%%%%%%%%%%%%
    
    \subsection{FAR(1) Model}
    \label{sec:FAR1}
        
        The most popular statistical model used to capture temporal dependence between functional observations is the functional autoregressive process.
        The theory of functional autoregressive processes in Hilbert spaces is developed in the pioneering monograph of \citealt{bosq} and a comprehensive collection of statistical advancements for the FAR model can be found in \citealt{horvath2012inference}. 

        A sequence of mean zero random functions $\{Y_t\}_{t =1}^{T} \subset \mathbb{H}$ follows a non-concurrent Functional Autoregressive Process of order 1 if:
        \begin{equation}
            \label{eq:FAR_equation}
            Y_{t} = \Psi Y_{t-1} + \varepsilon_{t} \qquad t=2,\dots,T, 
        \end{equation}
        where $\{ \varepsilon_{t} \}_{t \in \mathbb{N}}$ is a sequence of iid mean-zero innovation errors with values in $\mathbb{H}$ satisfying $\mathbb{E}[||\varepsilon_{t}||^2] < +\infty$ and $\Psi$ is a linear bounded operator from $\mathbb{H}$ to $\mathbb{H}$. We consider $\Psi$ to be a Hilbert-Schmidt operator with kernel $\psi$, in such a way that:
        \begin{equation}
            (\Psi x) (u,v) = \int_c^d \int_e^f \psi(u,v;w,z) x(w,z) dw dz \quad a.e., \, \forall (u,v) \in [c,d] \times [e,f].
        \end{equation}
        In order to ensure existence of a stationary solution of \eqref{eq:FAR_equation}, one has to require that $\exists j_0 \in \mathbb{N}$ such that $||\Psi||^{j_0} < 1$ (\citealt{bosq}, Lemma 3.1). 
        
        %---------------------------------------------------------
        
    \subsection{FAR(1) Estimation}
    \label{sec:FAR1_estimation}
        
        %Following a procedure similar to the Yule-Walker estimation in the scalar setting, we propose now a more sophisticated estimator of $\Psi$. Proceeding similarly to \citealt{horvath2012inference}, we derive the following estimator:
        Proceeding similarly to \citealt{horvath2012inference}, and adopting an approach akin to the Yule-Walker estimation in the scalar setting, we propose the following estimator of $\Psi$:        
        \begin{equation}
        \label{eq:EK_estimator}
            \hat{\Psi}_{M} x = \frac{1}{T-1} \sum_{i,j = 1}^{M} \sum_{t=1}^{T} \hat{\lambda}_j \langle x, \hat{\xi}_j \rangle \langle Y_{t-1}, \hat{\xi}_j \rangle \langle Y_{t}, \hat{\xi}_i \rangle \hat{\xi}_i
            \qquad a.e.,
        \end{equation}
        where $\xi_1, \dots ,\xi_M$ are the first M normalized functional principal components (FPC's), $\lambda_1, \dots, \lambda_M$ are the corresponding eigenvalues, and $\langle x, \xi_1 \rangle, \dots, \langle x, \xi_M \rangle$ are the scores of $x$ along the FPC's.
        \ref{appendix_FPCA} illustrates two different estimation techniques for $\xi_i$ and $\lambda_i$, one based on a discretization of functions on a fine grid and the other designed starting from an expansion of data on  a finite basis system.
        We further refer to \ref{appendix_EK} for more details on the derivation of estimator \eqref{eq:EK_estimator} and for an extensive discussion on how to adapt it to the Conformal Prediction setting, where $\Psi$ is estimated
        %the point predictor $g_{\mathcal{I}_1}$ 
        from the training set $\mathcal{I}_1$ only. 
        
        Another forecasting procedure based on FPC's has been proposed by \citealt{Aue} for one-dimensional functional data and is here extended to the two-dimensional setting.
        Calling once again $\xi_1, \dots, \xi_M$ the first $M$ functional principal components, we decompose the Functional Time Series as follows:
        \begin{align}
            Y_t(u,v) &= \sum_{j=1}^{M} \langle Y_t, \xi_j \rangle \xi_j(u,v) + e_t(u,v) = \bm{Y}_t^T \bm{\xi}(u,v) + e_t(u,v),
        \end{align}
        where $\bm{Y}_t=[\langle Y_t, \xi_1 \rangle, \dots, \langle Y_t, \xi_M \rangle]^T$ contains the projection scores, $\bm{\xi}(u,v) = [\xi_1(u,v),\dots,\xi_M(u,v)]^T$ collects the evaluated principal components, and $e_t(u,v)$ is the approximation error due to the expansion's truncation on the first $M$ principal components.
        Neglecting the approximation error $e_t$, one can prove that the vector $\bm{Y}_t$ follows a multivariate autoregressive process of order 1 (VAR(1)).
        Plugging in the estimated FPCs $\hat{\xi}_1, \dots, \hat{\xi}_M$, we can estimate the parameters of the resulting VAR(1) model using standard techniques of multivariate statistics and forecast $\hat{\bm{Y}}_{T+1}$ based on historical data $\bm{Y}_1,\dots,\bm{Y}_T$.
        The predicted function $\hat{Y}_{T+1}$ is then reconstructed as: 
        \begin{equation}
        \label{eqn:VAR_efpc_hat}
            \hat{Y}_{T+1}(u,v) = \hat{\bm{Y}}_{T+1}^T\bm{\xi}(u,v).
        \end{equation}
        
        We finally introduce a model that may appear simplistic, since it does not exploit the possible time dependence between functions' values in different points of the domain, but that in practical applications provides satisfying results. 
        The prediction method assumes an autoregressive structure in each location $(u,v)$ of the domain, ignoring the dependencies between different points. We call this model a \textit{concurrent} FAR(1):
        \begin{equation}
        \label{eqn:FAR_concurrent}
            Y_{t}(u,v) = \psi_{u,v} Y_{t-1}(u,v) + \varepsilon_{t}(u,v) \qquad \forall (u,v) \in [c,d] \times [e,f],
        \end{equation}
        where $\psi_{u,v} \in \mathbb{R}$ and $t=2,\dots,T$.
        Supposing to have observed functional data $y_1, \dots, y_T$ on a common two-dimensional grid $\{(u_i,v_j), i=1,\dots,N_1, j=1,\dots,N_2 \}$, 
        we can estimate $\psi_{u_i,v_j}$ for each location $(u_i,v_j)$.
        
        % Throughout the rest of the work, we will employ all the estimators presented above, comparing them in Section \ref{section:4} in terms of prediction performances.
        
%%%%%%%%%%%%%%%%%%%%%%%%%%%%%%%%%%%%%%%%%%%%%%%%%%%%%%%%%%%%%%

\section{Simulation study}
\label{section:4}
    
    \subsection{Study Design}
    The goal of this section is twofold: we aim to assess the quality of the proposed CP bands and evaluate different point predictors in terms of the resulting prediction regions. 
    Since this work is focused on uncertainty quantification, we compare forecasting performances by means of the resulting Conformal Prediction bands. 
    Firstly and foremost, we estimate the unconditional coverage by computing the \textit{empirical unconditional coverage} in order to compare it with the nominal confidence level $1-\alpha$. 
    In the second place, we consider the size of the prediction bands, since a small prediction region is preferable as it includes subregions of the sample space where the probability mass is highly concentrated (\citealt{Lei1}) and it is typically more informative in practical applications.
    %We stress the fact that the size of a prediction band should be evaluated only after veryfing that the method which outputted that specific prediction set guarantees the desired coverage, which represents the primary aspect when assessing prediction sets.

    We employ as a data generating process a FAR(1) model in order to evaluate the estimation routines presented in Section \ref{sec:FAR1}. 
    In order to benchmark forecasting performances, we examine the forecasting methods against a naive one: $\hat{Y}_{T+1}=Y_T$.
    By including a forecasting algorithm that is not coherent with the data generating process, we can illustrate how the presented CP procedure performs when a good point predictor $g_{\mathcal{I}_1}$ is not available. Although as reported in Section \ref{section:point_prediction} a sufficiently accurate forecasting algorithm is necessary to guarantee asymptotic validity, we notice that in the simulations CP bands remain valid even when such assumption is not met.
    %Secondly, we explore beyond the FAR(1), simulating data from a Functional Moving Average (FMA) process. The aim of this last study is simply to asses the forecasting performances of the different point predictors presented in \ref{section:4} in a general case, where the data generating process does not comply with any point predictor.
    
    %In each scenario, we compare the performances of five selected prediction algorithms, one of which do not exploit the autoregressive structure. To obtain further insights, we also include the errors obtained by assuming perfect knowledge of the operator $\Psi$. For ease of reference, we briefly describe these methods, and introduce some convenient notation.
    
    To obtain further insights, we include the performances obtained by assuming perfect knowledge of the operator $\Psi$. For ease of reference, we list here the forecasting algorithms, introducing some convenient notation.
    
    \begin{itemize}
        \item \textbf{FAR(1)-Concurrent} refers to the forecasting algorithm based on a concurrent FAR(1) model \eqref{eqn:FAR_concurrent}.
        \item \textbf{FAR(1)-EK} (Estimated Kernel) denotes the first estimation procedure presented in Section \ref{sec:FAR1}, where we explicitly compute $\hat{\Psi}_M$ as prescribed by \eqref{eq:EK_estimator} and then set $\hat{Y}_{T+1} = \hat{\Psi}_M Y_T$.
        \item \textbf{FAR(1)-EK+} (Estimated Kernel Improved) is a modification of the above method, where eigenvalues $\hat{\lambda}_i$ are replaced by $\hat{\lambda}_i + 1.5(\hat{\lambda}_1+\hat{\lambda}_2)$, as recommended by \citealt{didericksen} and discussed in \ref{appendix_EK}.
        \item \textbf{FAR(1)-VAR} denotes the forecasting procedure \eqref{eqn:VAR_efpc_hat},
        where we exploit the expansion on estimated functional principal components and forecast $Y_{T+1}$ using the underlying VAR(1) model.
        \item \textbf{Naive}: we just set $\hat{Y}_{T+1}=Y_T$. This method does not attempt to model temporal evolution, it is only included to see how much can be gained by exploiting the autoregressive structure of data.
        \item \textbf{Oracle}: we set $Y_{T+1}=\Psi Y_T$, using the same exact operator $\Psi$ from which data are simulated. 
        This point predictor is clearly not available in practical application, but it is interesting to include it in order to see if poor predictions might be due to poor estimation of $\Psi$.
    \end{itemize}
    
    When it is required (namely in FAR(1)-EK, FAR(1)-EK+, FAR(1)-VAR), FPCA is performed using the discretization approach, as motivated in \ref{appendix_FPCA}, truncating the representation to the first 4 harmonics.
    %The number of principal components is selected by the cumulative proportion of variance criterion.
    %Calling $\hat{\lambda}_1,\dots,\hat{\lambda}_M$ the $M$ largest estimated eigenvalues, we choose $M\in \mathbb{N}$ such that $\sum_{j=1}^{M}\hat{\lambda}_j / \sum_{j=1}^{\infty}\hat{\lambda}_j$ exceeds a predetermined percentage value, in this case fixed equal to 0.8. We noticed that, on average, this entails to select a number of harmonics equal to 5. 
    
    %Throughout the whole simulation study, we set the significance level $\alpha=0.1$.
    In Section \ref{subsec:sample_size}, we  fix the size $b$ of the blocking scheme \eqref{eqn:blocking_scheme} equal to 1 and let the sample size $T$ take values $19,49,99,499$. Secondly, in Section \ref{subsec:block_size}, we instead fix the sample size equal to $119$ and repeat the simulations with $b=1,3,6$. As usually done in the time series setting, the first observation is taken into account as a covariate only and does not enter neither the training set nor the calibration set. The proportion of data in the training and in the calibration set are hence equal to one half of the remaining observations: $m=l=(T-1)/2$. 
    %Thanks to the chosen values of $T$, $l$ and $\alpha$, we can guarantee an actual coverage of $1-\frac{\left \lfloor{(l+1)\alpha}\right \rfloor}{(l+1)} = 1-\alpha$. 
    For each value of $T$, we repeat the procedure by considering $N=1000$ simulations. Simulations are implemented in the R Programming Language (\citealt{R}).
    
\begin{figure}
        \centering
        \includegraphics[width=0.8\linewidth]{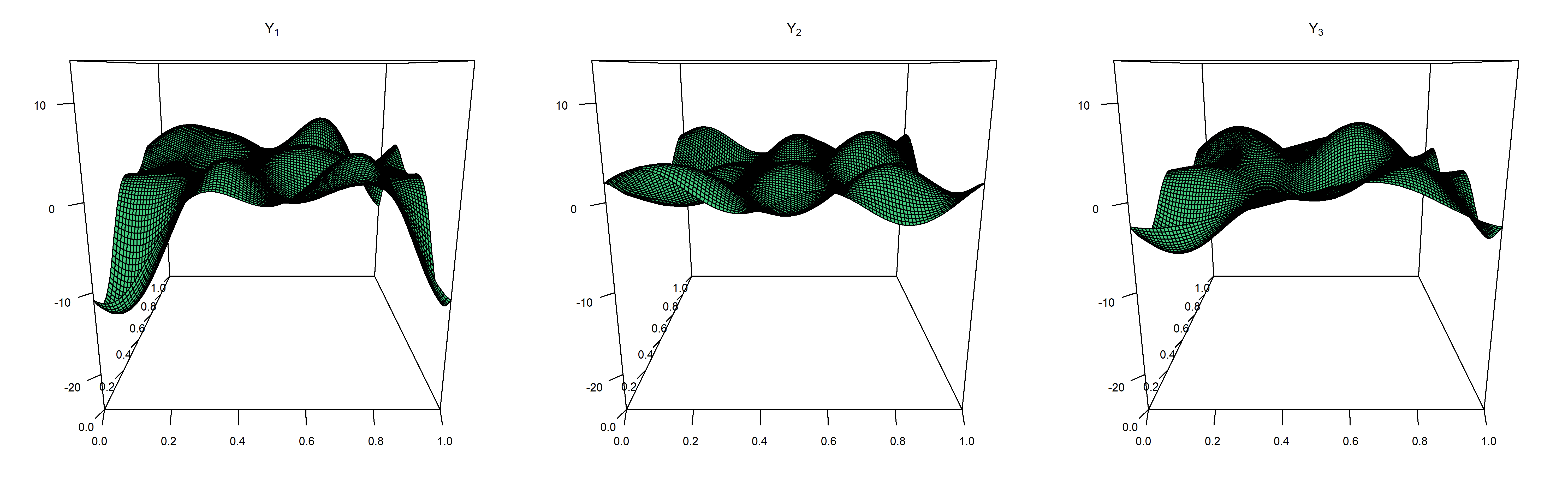}
        \caption[Representation of a simulated FAR(1)]{Three consecutive realizations of a simulated FAR(1) represented on a grid of $10^4$ points.}
        \label{fig:evolution}
    \end{figure}

    %In order to simulate a sequence of functions $\{Y_t\}_{t=1,\dots,T}$ from a Functional Autoregressive Process of order one, we extend the implementation of \citealt{freqdomftsa} and assume that observations lie in a finite dimensional subspace of the function space $\mathbb{H}$, spanned by orthonormal basis functions $\phi_1, \dots, \phi_M$, with $M\in \mathbb{N}$ representing the dimension of such subspace. 
    %Similarly to \citealt{freqdomftsa}, 
    In order to simulate a sequence of functions $\{Y_t\}_{t=1,\dots,T}$ from a FAR(1), we assume that observations lie in a finite dimensional subspace of the function space $\mathbb{H}$. Without loss of generality, throughout this section we consider functions in $\mathbb{H}=\mathcal{L}^2([0,1]\times[0,1])$. $\mathbb{H}$ is spanned by orthonormal basis functions $\phi_1, \dots, \phi_M$, with $M\in \mathbb{N}$ representing the dimension of such subspace. 
    Therefore, we have:
    \begin{gather}
        Y_{t}(u,v) = \bm{\phi}(u,v)^T \bm{Y}_t, \\
        \varepsilon(u,v) = \bm{\phi}(u,v)^T \bm{\varepsilon}_t, \\
        \label{eqn:kernel_basis}
        \psi(u,v;w,z) = \bm{\phi}(u,v)^T \bm{\Psi} \bm{W} \bm{\phi}(w,z),
    \end{gather}
    where $\bm{\phi}(u,v)=[\phi_1(u,v), \dots, \phi_M(u,v)]^T \in \mathbb{R}^M, \, \forall (u,v) \in [0,1]\times[0,1]$, $\bm{Y}_t, \,\bm{\varepsilon}_t \subset \mathbb{R}^M, \, \forall t = 1, \dots, M$ and $\bm{\Psi} \in \mathbb{R}^{M \times M}$ and $\bm{W} \in \mathbb{R}^{K \times K}$, defined as $\bm{W} := \int_c^d \int_e^f \bm{\phi}(u,v) \bm{\phi}(u,v)^T du dv$.
    It follows that:
    \begin{equation}
        \bm{Y}_{t} = \bm{\Psi} \bm{Y}_{t-1} \bm{W} + \bm{\varepsilon}_t \qquad t=1,\dots,T.
    \end{equation}
    The basis system $\phi_1, \dots, \phi_M$ is constructed as the tensor product basis of two cubic B-spline systems $\{g_i\}_{i = 1,\dots, M_1}$, $\{h_j\}_{j = 1,\dots, M_2}$, both defined on $[0,1]$. We set $M_1=M_2=5$, so that $M=25$. 
    %Notice that, by including more functions, we can better approximate the space $\mathbb{H}$, though inevitably producing rougher curves. On the other hand, by reducing the size of the basis system, one renounce to have a good representation of $\mathbb{H}$, but this permits to obtain smoother functions. The choice proposed for $M$ is arbitrary, but provides a good compromise between the two outlined extremes. 
    For a discussion on the tensor product basis system, we refer to \ref{sec:FPCA_basis}
    The matrix $\bm{\Psi}$ is defined as $\bm{\Psi} :=0.7 \frac{\tilde{\bm{\Psi}}}{||\tilde{\bm{\Psi}}||_F}$, with $\tilde{\bm{\Psi}}$ having diagonal values equal to $0.8$ and out-diagonal elements equal to 0.3
    Innovation errors $\bm{\varepsilon}_t$ are independently sampled from a multivariate 
    %Student's $t$-distribution, with 4 degrees of freedom and scale matrix %$\bm{\Sigma}$ having diagonal elements equal to 0.5 and out-diagonal entries equal to 0.3. 
    normal distribution, with mean zero and covariance matrix $\bm{\Sigma}$ having diagonal elements equal to 0.5 and out-diagonal entries equal to 0.3. 
    \autoref{fig:evolution} depicts an example of a simulated Functional Autoregressive Process of order one. A GIF of the time-evolving FAR(1) process can be found on \href{https://github.com/Niccolo-Ajroldi/ARMA-Surfaces/blob/main/Pics/FAR.gif}{GitHub}.

    Notice that simulations have been designed in such a way to generate a \textit{stationary process}. This condition is important to guarantee the existence of a solution to the FAR(1) equation \eqref{eq:FAR_equation}, and is here guaranteed by setting $||\Psi|| < 1$, which satisfies the sufficient condition for stationary presented in Lemma 3.1 of \citealt{bosq}. 
    One can indeed prove that, if relation \eqref{eqn:kernel_basis} holds, then $||\Psi||=||\bm{\Psi}||_F$, where $||.||$ is the usual operatorial norm and $||.||_F$ denotes the Frobenius norm.
    In this way, FAR(1) estimation techniques are  well-defined.
    For what concerns the theoretical assumptions of the CP scheme, proving the hypothesis of Theorem \autoref{my_theorem} is difficult, in particular in the context of functional data. To the best of our knowledge, no test for strongly mixing two-dimensional time series has been proposed, and testing the bounds on the oracle nonconformity scores is even more challenging.
    However, we aim to show here how the CP procedure can still be applied to obtain valid and efficient prediction bands.
    
    %%%%%%%%%%%%%%%%%%%%%%%%%%%%%%%%%%%%%%%%%%%%%%%%%%%%%%%%%%%%%%%%%%%%%%%%%
    \subsection{Varying the the Sample Size}
    \label{subsec:sample_size}
    
    We first fix the size $b$ of the blocking scheme equal to 1 and let the sample size $T$ take values $19,49,99,499$. We replicate the experiments with different significance levels: $\alpha=0.1$, $\alpha=0.2$ and $\alpha=0.3$, in order to assess how the confidence level influences the coverage and the width of the prediction bands.
    
\begin{figure}[]
        \centering
        \begin{subfigure}{.33\textwidth}
            \centering \includegraphics[width=1\linewidth]{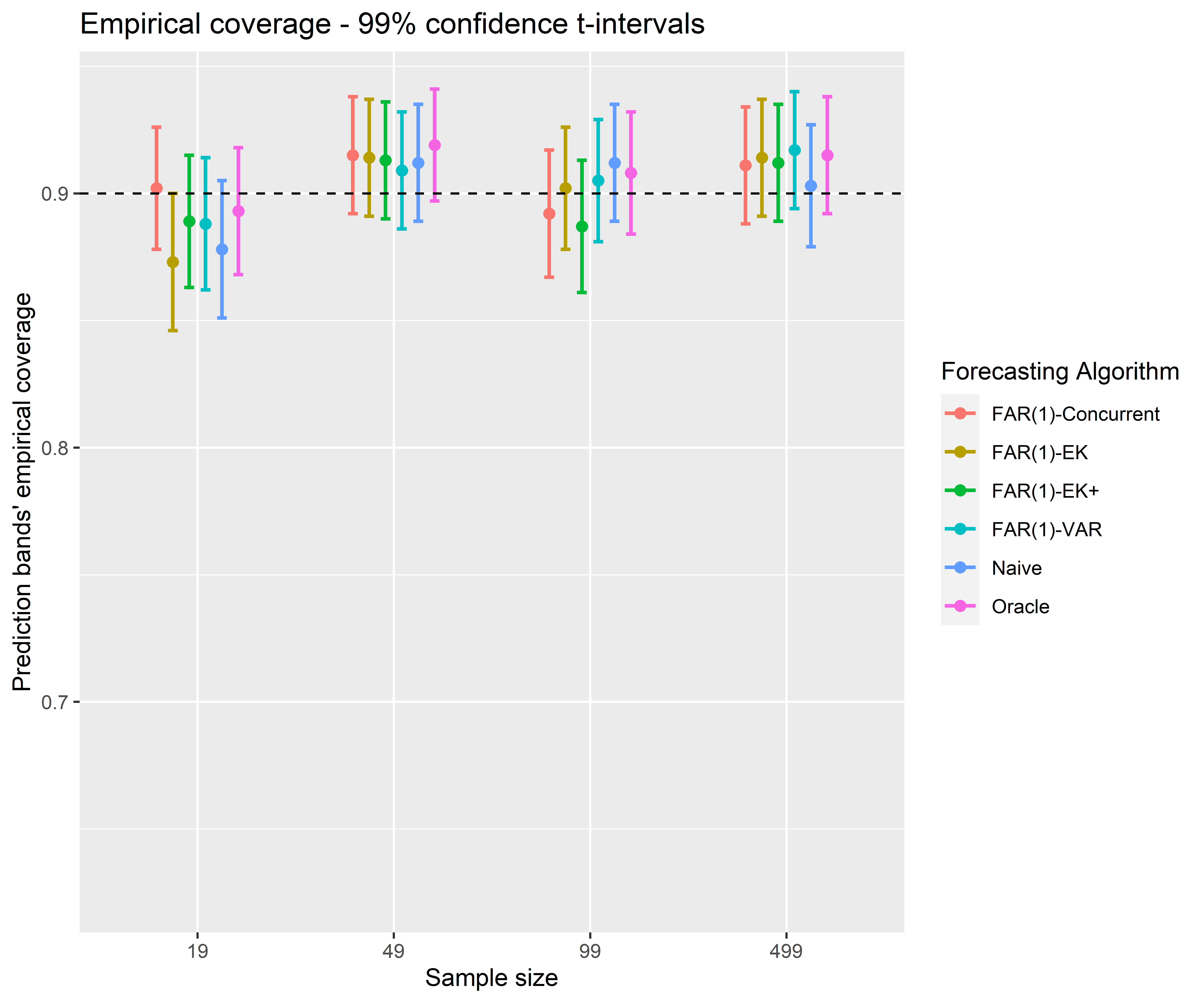}
            \caption{$\alpha=0.1$}
            \label{fig:sim_1_alpha_0_1_coverage}
        \end{subfigure}%
        \begin{subfigure}{.33\textwidth}
            \centering \includegraphics[width=1\linewidth]{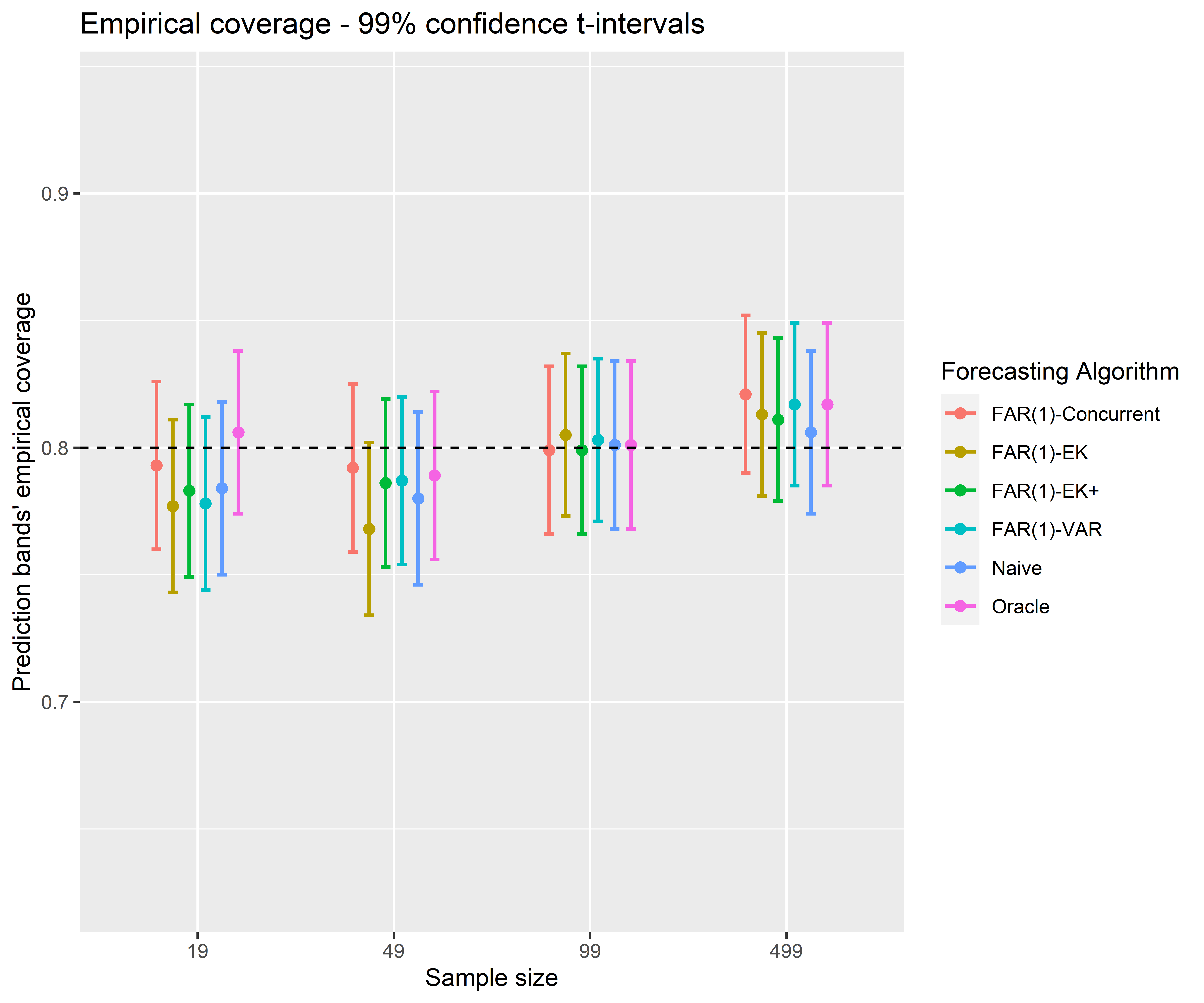}
            \caption{$\alpha=0.2$}
            \label{fig:sim_1_alpha_0_2_coverage}
        \end{subfigure}
        \begin{subfigure}{.33\textwidth}
            \centering \includegraphics[width=1\linewidth]{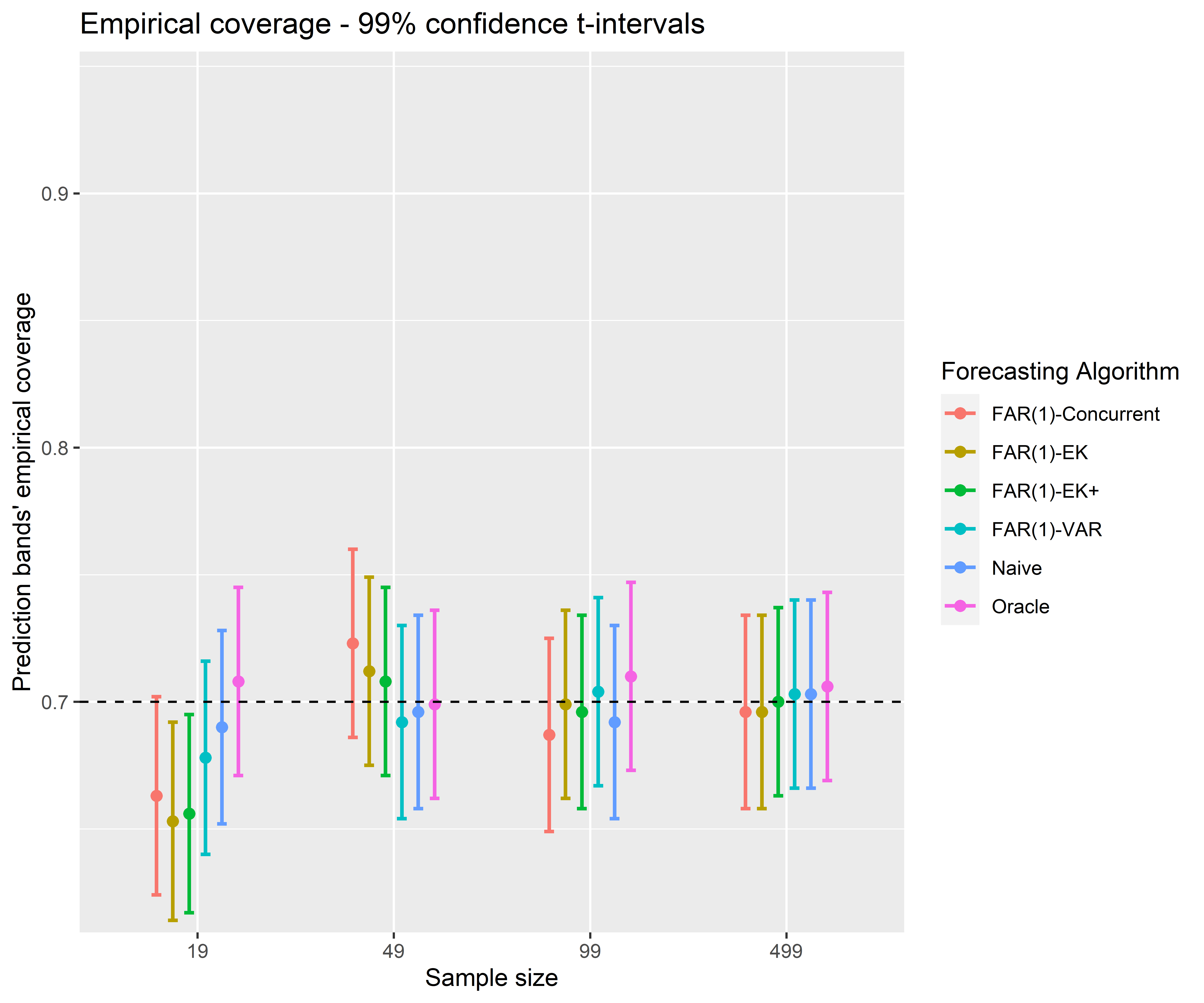}
            \caption{$\alpha=0.3$}
            \label{fig:sim_1_alpha_0_3_coverage}
        \end{subfigure}%
    \caption{Simulation study, increasing the sample size. Empirical coverage of CP bands. Dashed line represents nominal coverage $1-\alpha$.}
    \label{fig:sim_1_coverage}
    \end{figure}

\begin{figure}[]
        \centering
        \begin{subfigure}{.33\textwidth}
            \centering \includegraphics[width=1\linewidth]{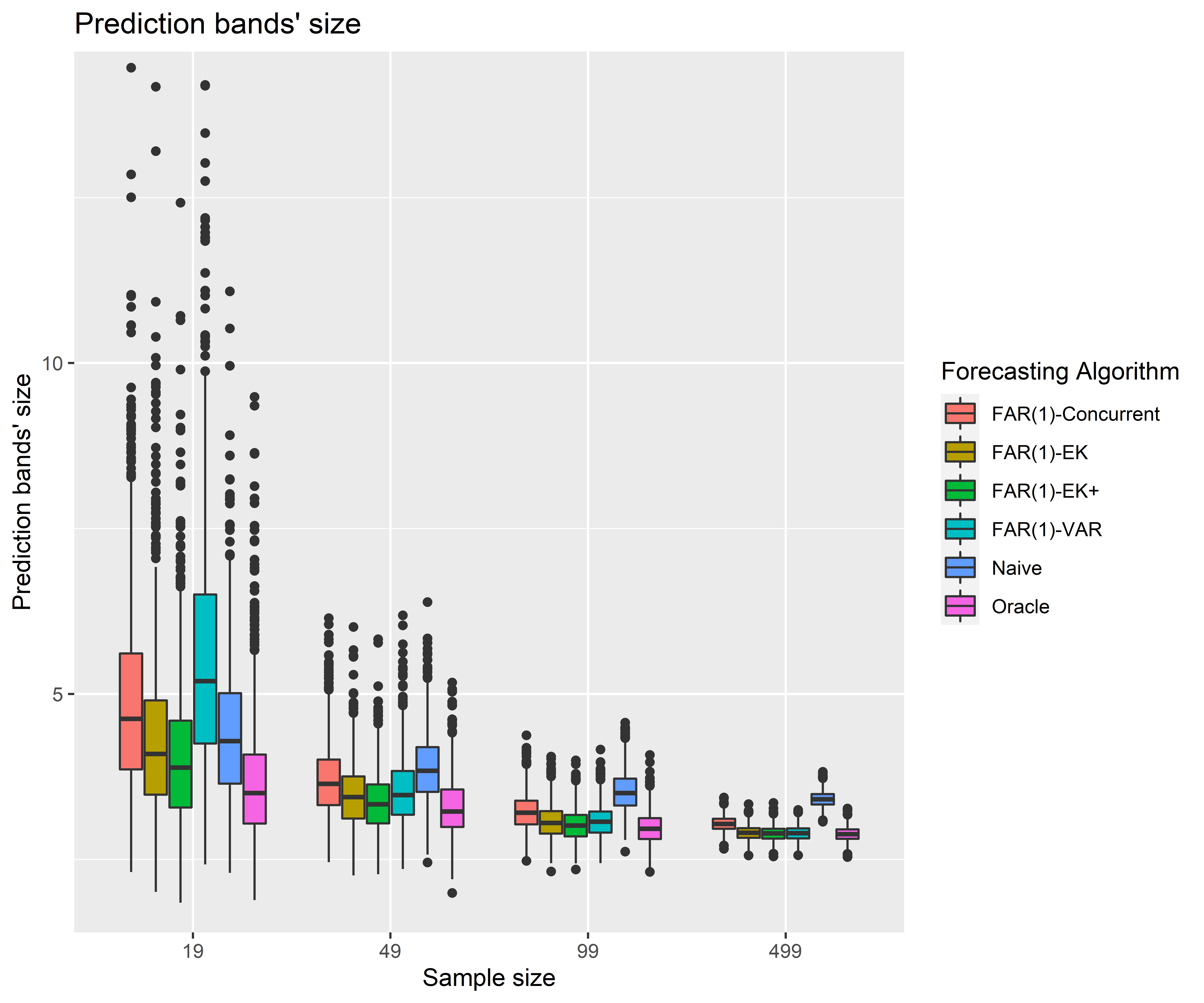}
            \caption{$\alpha=0.1$} 
            \label{fig:sim_1_alpha_0_1_width}
        \end{subfigure}%
        \begin{subfigure}{.33\textwidth}
            \centering \includegraphics[width=1\linewidth]{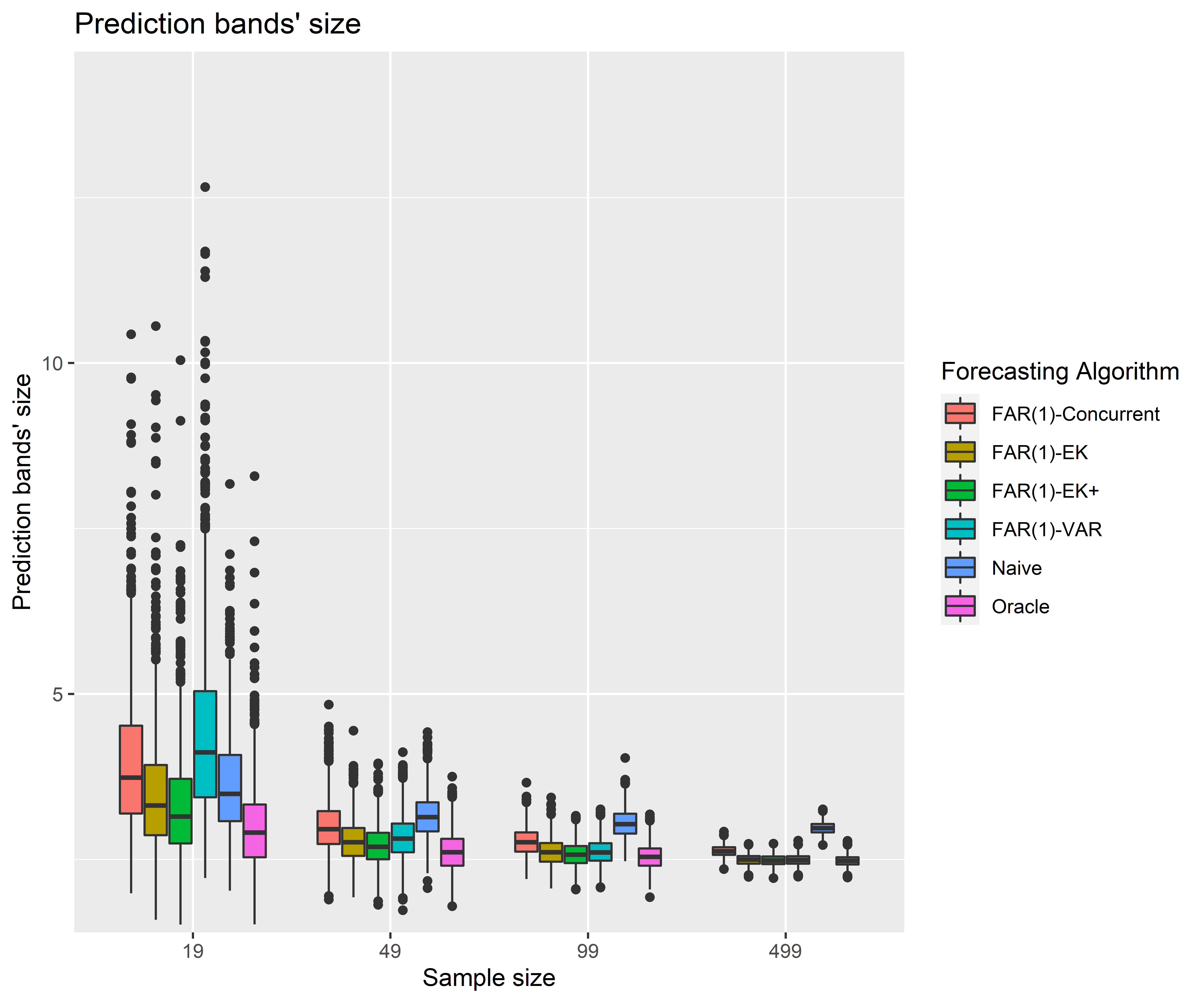}
            \caption{$\alpha=0.2$}
            \label{fig:sim_1_alpha_0_2_width}
        \end{subfigure}
        \begin{subfigure}{.33\textwidth}
            \centering \includegraphics[width=1\linewidth]{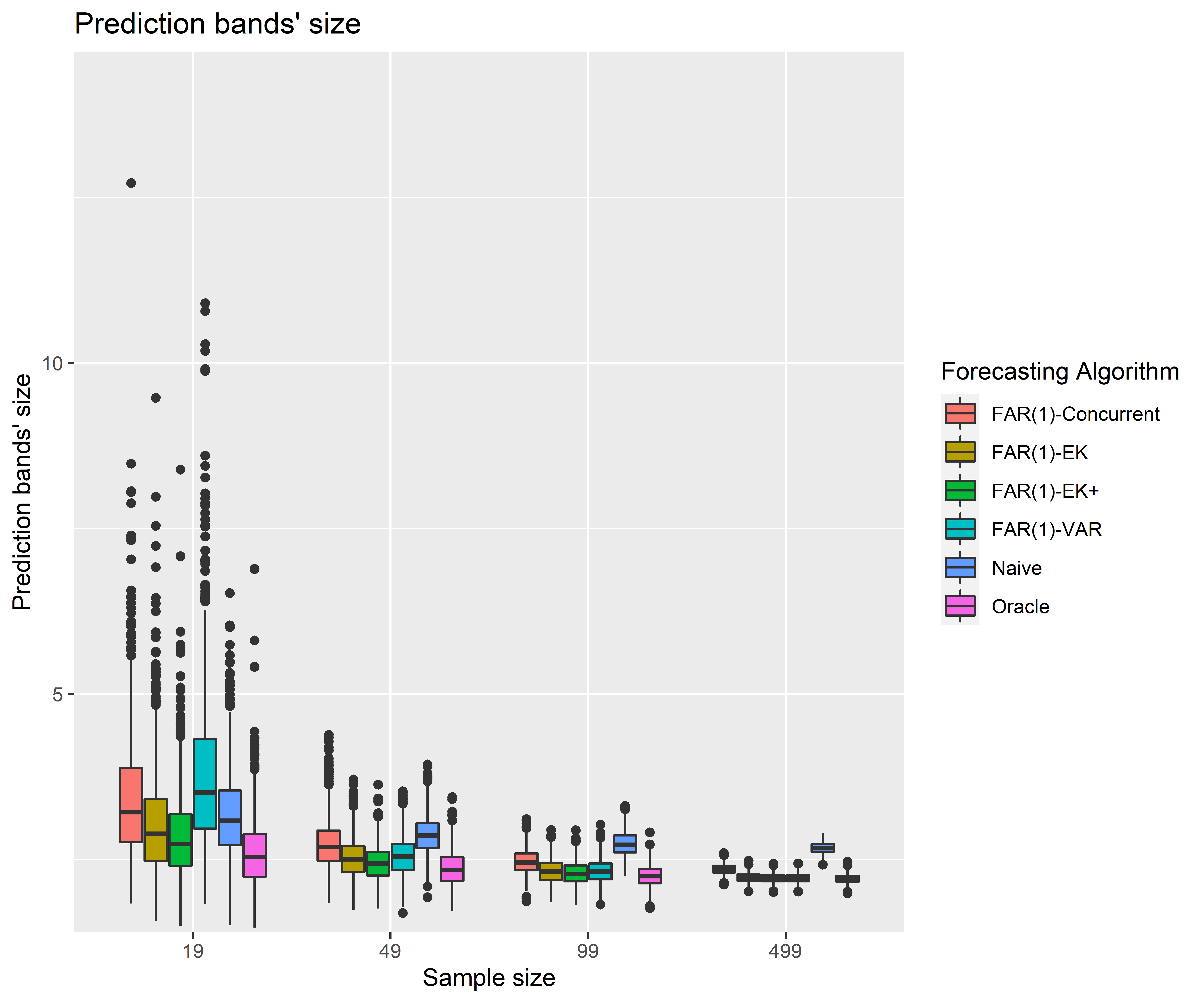}
            \caption{$\alpha=0.3$}
            \label{fig:sim_1_alpha_0_3_width}
        \end{subfigure}%
    \caption{Simulation study, increasing the sample size. Size of CP bands.}
    \label{fig:sim_1_width}
    \end{figure}

    % \autoref{fig:sim_1_alpha_0_1_coverage}, \autoref{fig:sim_1_alpha_0_2_coverage} and \autoref{fig:sim_1_alpha_0_3_coverage} 
    \autoref{fig:sim_1_coverage} shows the empirical coverage, together with the related 99\% confidence interval. 
    Empirical coverage is computed as the fraction of the $N=1000$ replications in which $y_{T+1}$ belongs to $\mathcal{C}_{T,1-\alpha}(x_{T+1})$, and the confidence interval is reported in order to provide insights into the variability of the results, rather than to draw inferential conclusions about the unconditional coverage. 
    Notice that different point predictors might intrinsically have dissimilar coverages, consequently this analysis aims to compare forecasting algorithm in terms of their predictive performances.
    We can appreciate that the 99\% confidence interval for the empirical coverage almost always includes the nominal confidence level, regardless of the sample size at disposal. The only exception is obtained with $\alpha=0.3$ and $T=19$. In this case, the method produces very narrow prediction bands (\autoref{fig:sim_1_alpha_0_3_width}), that result in an empirical coverage smaller than the nominal one. This behavior however disappears as soon as the sample size $T$ increases.
    It is also interesting to notice that, even when an accurate forecasting algorithm $g_{\mathcal{I}_1}$ is not available (namely with the Naive predictor), the proposed CP procedure still outputs prediction regions with a high unconditional coverage. 
    
    Similarly to \citealt{diquigiovanni2021importance}, we define the size of a two-dimensional prediction band as the \textit{volume} between the upper and the lower surfaces that define the prediction band:
    \begin{equation}
        \mathcal{Q}(s_{\mathcal{I}_1}) := \int_0^1 \int_0^1 2 k s_{\mathcal{I}_1}(u,v) du dv = 2 k.
    \end{equation}

    Measuring the size of the correspondent prediction bands, we can compare the efficiency of different forecasting routines.
    We stress the fact that distinct point predictors may guarantee potentially different coverage levels. For this reason, it is crucial to first evaluate the empirical coverage of the resulting prediction bands and only afterward investigate their size.
    \autoref{fig:sim_1_width}
    %\autoref{fig:sim_1_alpha_0_1_width}, \autoref{fig:sim_1_alpha_0_2_width} and \autoref{fig:sim_1_alpha_0_3_width} 
    reports boxplots with prediction bands' size for the $N=1000$ simulations and for different values of $\alpha$ and $T$.
    Bands' size tends to decrease as long as the number of observations $T$ increases, hence improving the efficiency of prediction sets. 
    Moreover, the size tends to decrease when the confidence level $1-\alpha$ increases.
    As expected, Naive predictor provides larger prediction bands, particularly in the large sample settings.
    %and when $T$ grows the size of the prediction regions of other methods systematically dominates the Naive's one. 
    %On the other hand, the forecasting algorithm based on a Concurrent FAR(1) model \eqref{eqn:FAR_concurrent} tends to produce narrower prediction bands with respect to the naive reference predictor, thanks to the local autoregressive structure assumed.
    On the other hand, FAR(1)-EK and FAR(1)-EK+, both based on the estimation of the autoregressive operator $\Psi$, provide the tightest prediction bands, not only when numerous observations are available, but also in small sample sizes scenario.
    Notice also that eigenvalue correction slightly improves the performances of FAR(1)-EK+ wrt FAR(1)-EK, especially when few samples are available.
    %Moreover, one can notice that FAR(1)-EK+ do not significantly improves FAR(1)-EK, neither in terms of coverage, nor in terms of band size. 
    %\footnote{The fact that such method do not improve...  has also been reported by \citealt{didericksen}, but we...}
    We acknowledge that, when $T=19$, VAR-efpc performs remarkably worse than the other methods. We argue that this performance gap might be caused by the simultaneous OLS estimation of the underling VAR(1) equations, which might provide biased estimates if the sample size is small.
    However, when the sample size increases, such forecasting algorithm performs comparably with the already mentioned FAR(1)-EK and FAR(1)-EK+.
    Finally, although the Conformal Prediction bands produced by the oracle predictor are obviously the most performing one, both FAR(1)-EK and FAR(1)-EK+ provide CP bands with coverage and size comparable to the theoretically perfect oracle forecasting method.
    
    %We repeat the experiment, varying the nominal coverage $\alpha$, reporting in ?? and ?? the results with $\alpha=0.05$ and $\alpha=$ respectively.
    %Once again, all the method achieve an empirical coverage very close to the nominal one, and the most narrow bands are produced by the methods ???

    %%%%%%%%%%%%%%%%%%%%%%%%%%%%%%%%%%%%%%%%%%%%%%%%%%%%

    \begin{figure}[h]
        \centering
        \begin{subfigure}{.33\textwidth}
            \centering \includegraphics[width=1\linewidth]{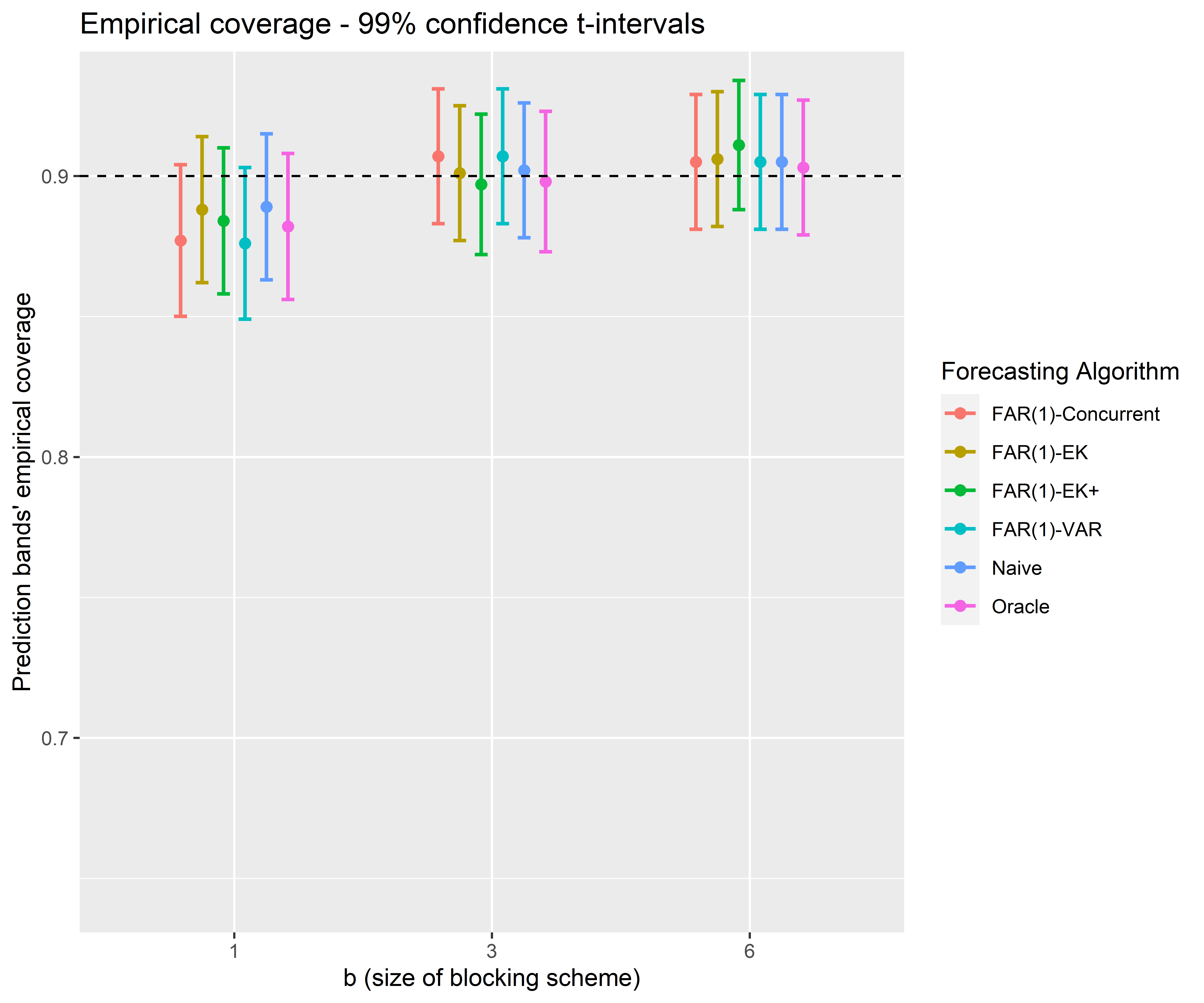}
            \caption{$\alpha=0.1$}
            \label{fig:sim_2_alpha_0_1_coverage}
        \end{subfigure}%
        \begin{subfigure}{.33\textwidth}
            \centering \includegraphics[width=1\linewidth]{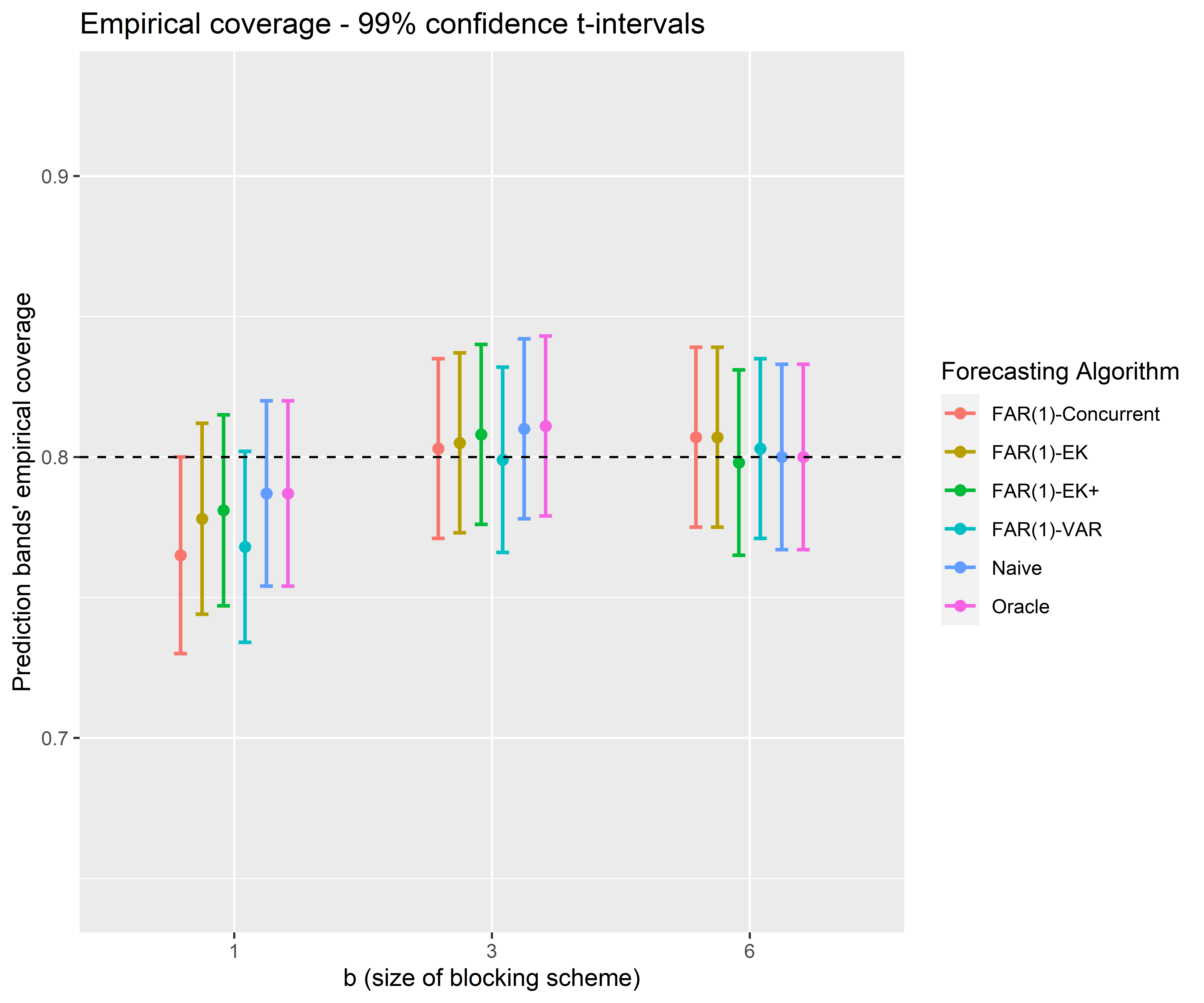}
            \caption{$\alpha=0.2$}
            \label{fig:sim_12_alpha_0_2_coverage}
        \end{subfigure}
        \begin{subfigure}{.33\textwidth}
            \centering \includegraphics[width=1\linewidth]{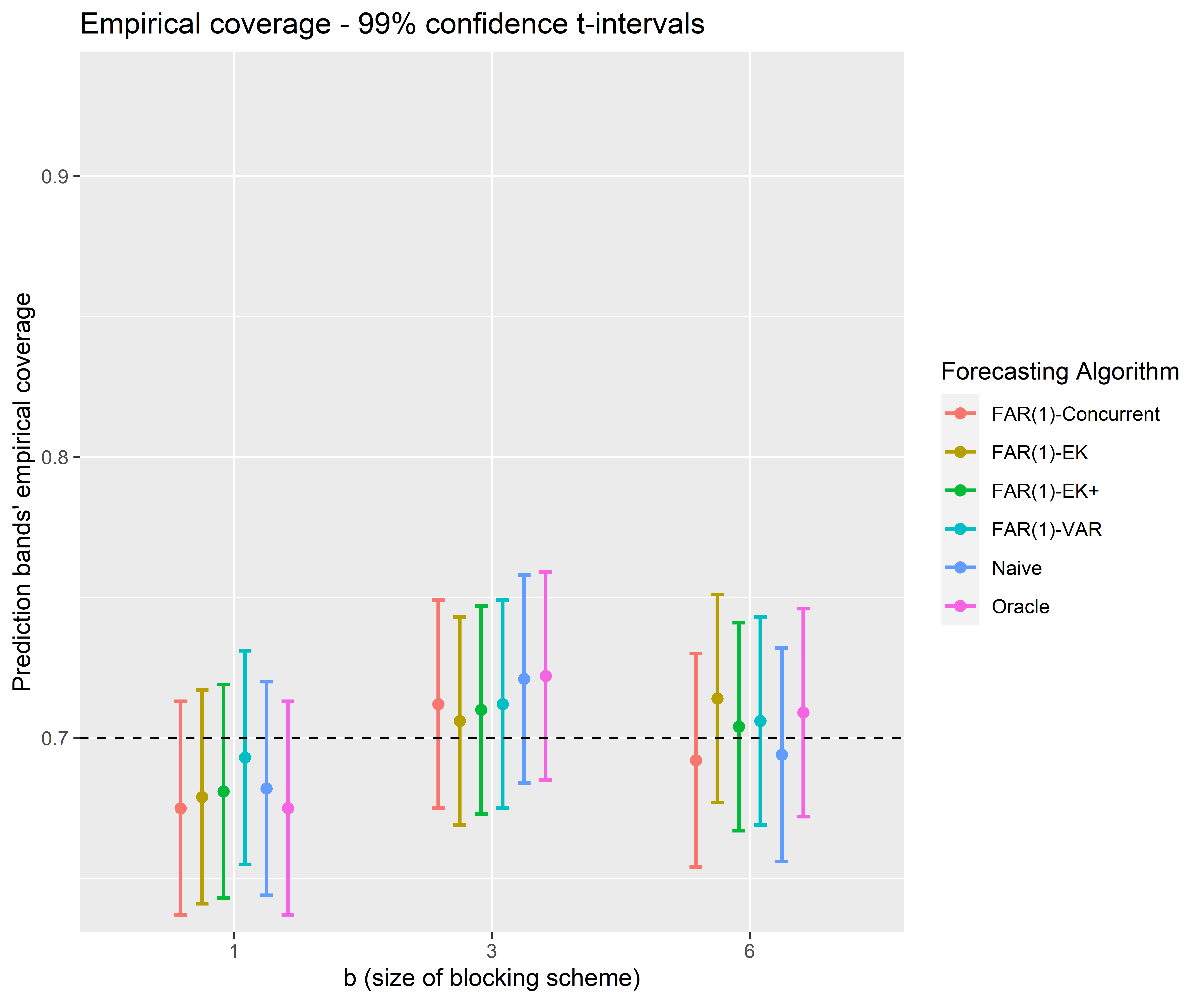}
            \caption{$\alpha=0.3$}
            \label{fig:sim_12_alpha_0_3_coverage}
        \end{subfigure}%
    \caption{Simulation study, increasing the blocking scheme size. Empirical coverage of CP bands. Dashed line represents nominal coverage $1-\alpha$.}
    \label{fig:sim_2_coverage}
    \end{figure}
    \begin{figure}[h]
        \centering
        \begin{subfigure}{.33\textwidth}
            \centering \includegraphics[width=1\linewidth]{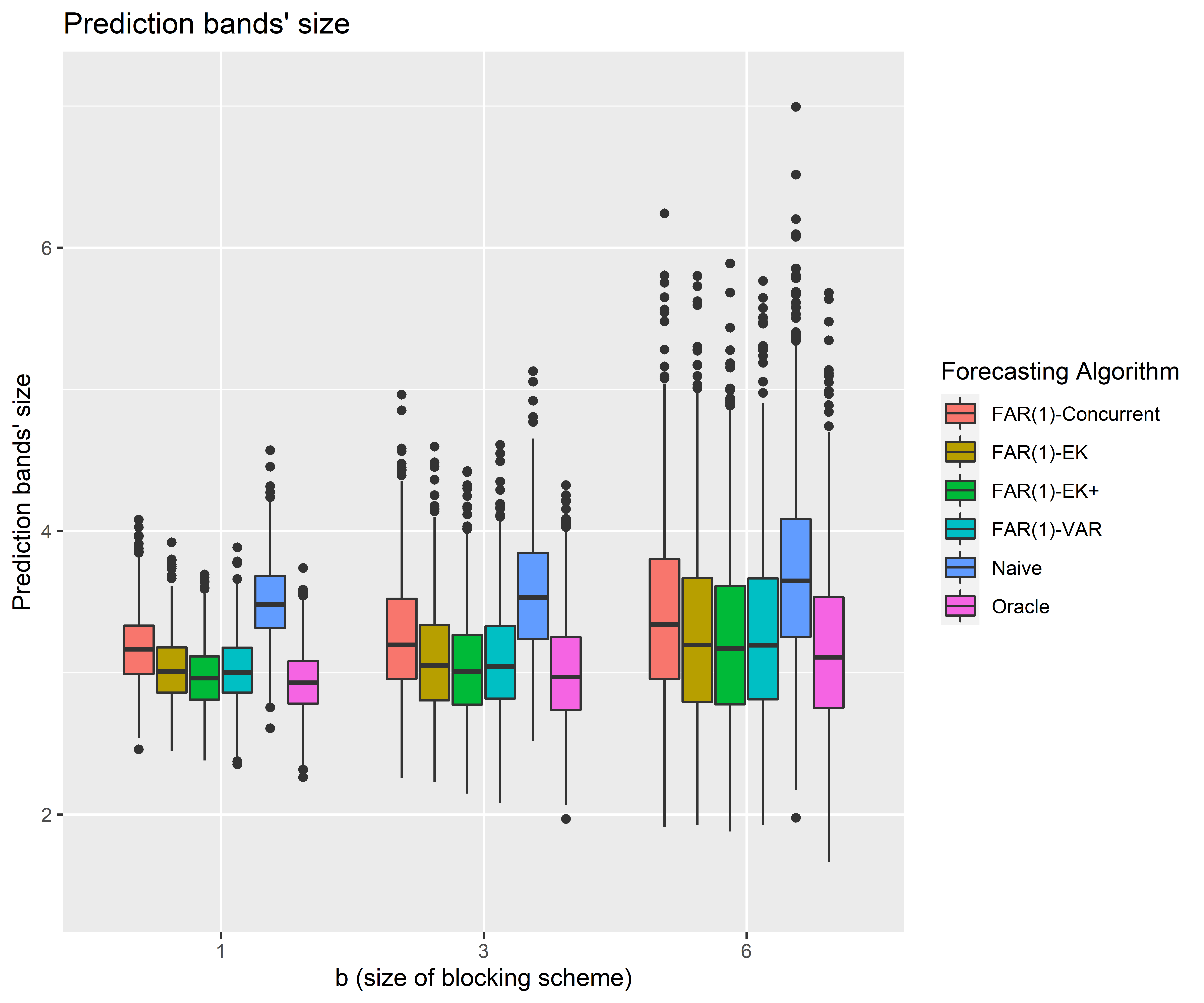}
            \caption{$\alpha=0.1$} 
            \label{fig:sim_2_alpha_0_1_width}
        \end{subfigure}%
        \begin{subfigure}{.33\textwidth}
            \centering \includegraphics[width=1\linewidth]{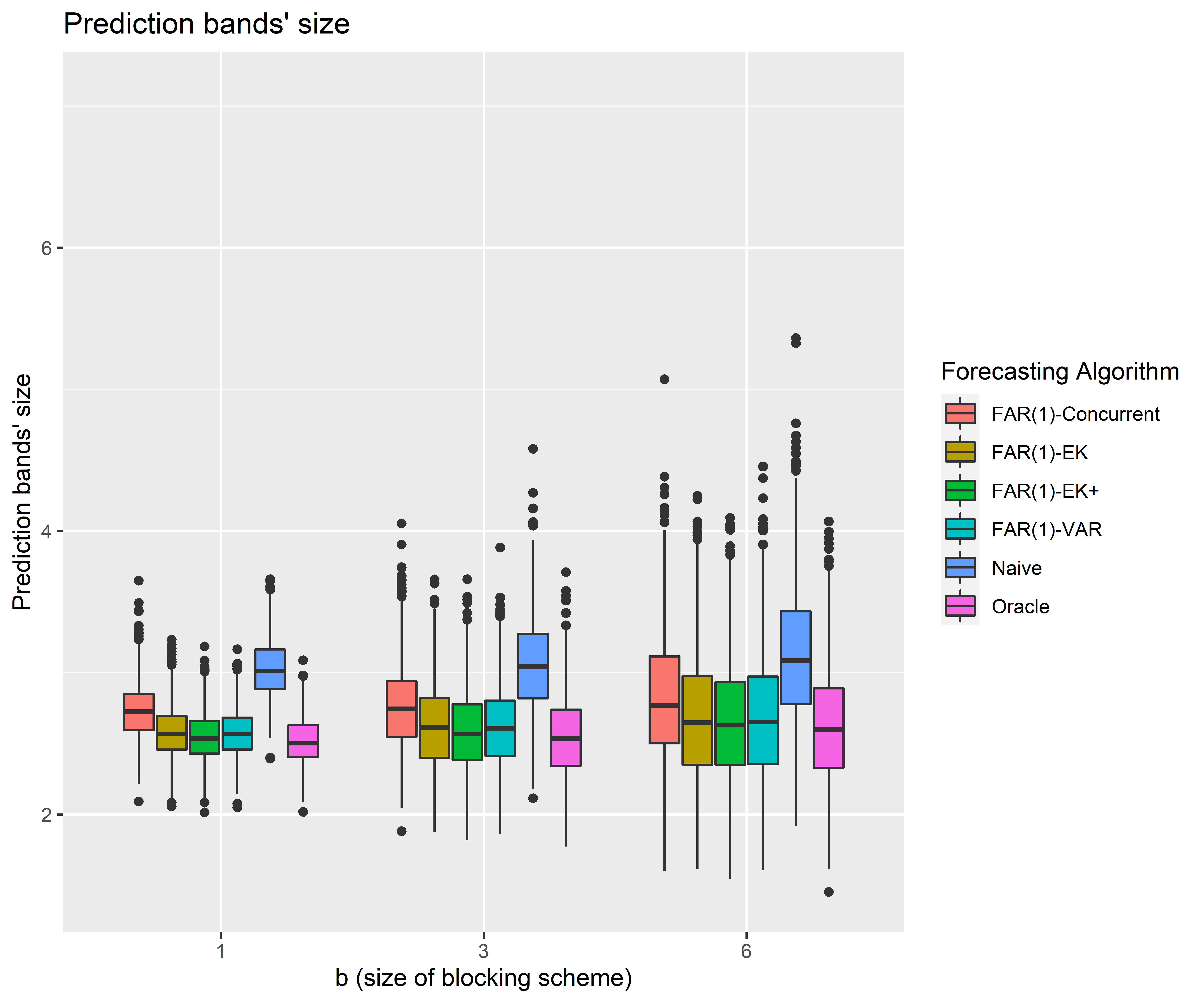}
            \caption{$\alpha=0.2$}
            \label{fig:sim_2_alpha_0_2_width}
        \end{subfigure}
        \begin{subfigure}{.33\textwidth}
            \centering \includegraphics[width=1\linewidth]{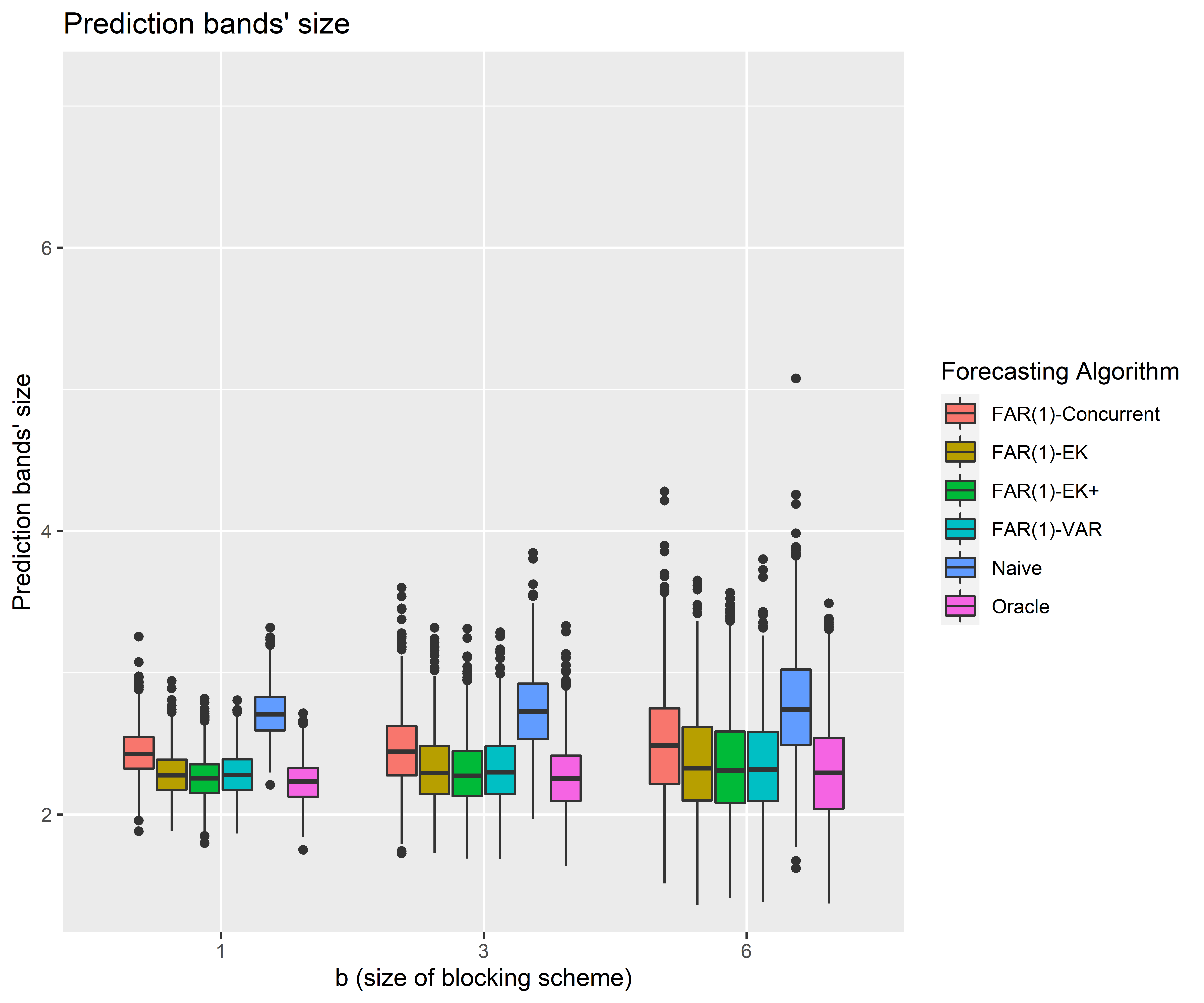}
            \caption{$\alpha=0.3$}
            \label{fig:sim_2_alpha_0_3_width}
        \end{subfigure}%
    \caption{Simulation study, increasing the blocking scheme size. Size of CP bands.}
    \label{fig:sim_2_width}
    \end{figure}

    \subsection{Varying the Blocking Scheme Size}
    \label{subsec:block_size}
    
    This time, we fix the sample size $T$ and let the blocking scheme $b$ vary, in order to determine how the validity and efficiency of the resulting prediction bands are influenced by such parameter. 
    Once again, we repeat the experiments for $\alpha=0.1$, $\alpha=0.2$, $\alpha=0.3$, whereas the value of $T$ is fixed and equal to 119, which provides a good balance between scenarios with small and large sample sizes. Analogous results have been found by letting the value of $T$ vary.
    Results are reported in \autoref{fig:sim_2_coverage} and \autoref{fig:sim_2_width}.
    Once again, in all circumstances the empirical coverage is close to the nominal one, confirming validity of CP bands even for higher values of $b$. 
    Moreover, one can notice that, as already pointed out by \citealt{diquigiovanni2021_FTS}, the band size tends to decreases when $b$ decreases, thus providing more efficient prediction regions.
    We argue that this behaviour is related to the inverse proportionality between the blocking scheme size $b$ and the dimension of permutation family $|\Pi|$. 
    % Indeed, recalling that $|\tilde{\Pi}|=\frac{l+1}{b}$, by employing lower value of $b$, the scheme makes use of a higher number of nonconformity scores, providing a smoother randomization p-value and hence a coverage closer to the nominal one.
    Finally, notice that larger prediction bands attain as expected a larger empirical coverage, and that larger values of $\alpha$ (smaller confidence level $1-\alpha$) result in smaller prediction bands.
    A comparison of forecasting algorithms performances validates the considerations in Section \ref{subsec:sample_size}.

%%%%%%%%%%%%%%%%%%%%%%%%%%%%%%%%%%%%%%%%%%%%%%%%%%%%%%%%%%%%%%

\section{Case Study: Forecasting Black Sea Level Anomalies}
\label{section:BlackSea}

\subsection{Dataset}
\begin{figure}[]
        \centering
        \begin{subfigure}{.5\textwidth}
            \centering 
            \includegraphics[width=.9\linewidth]{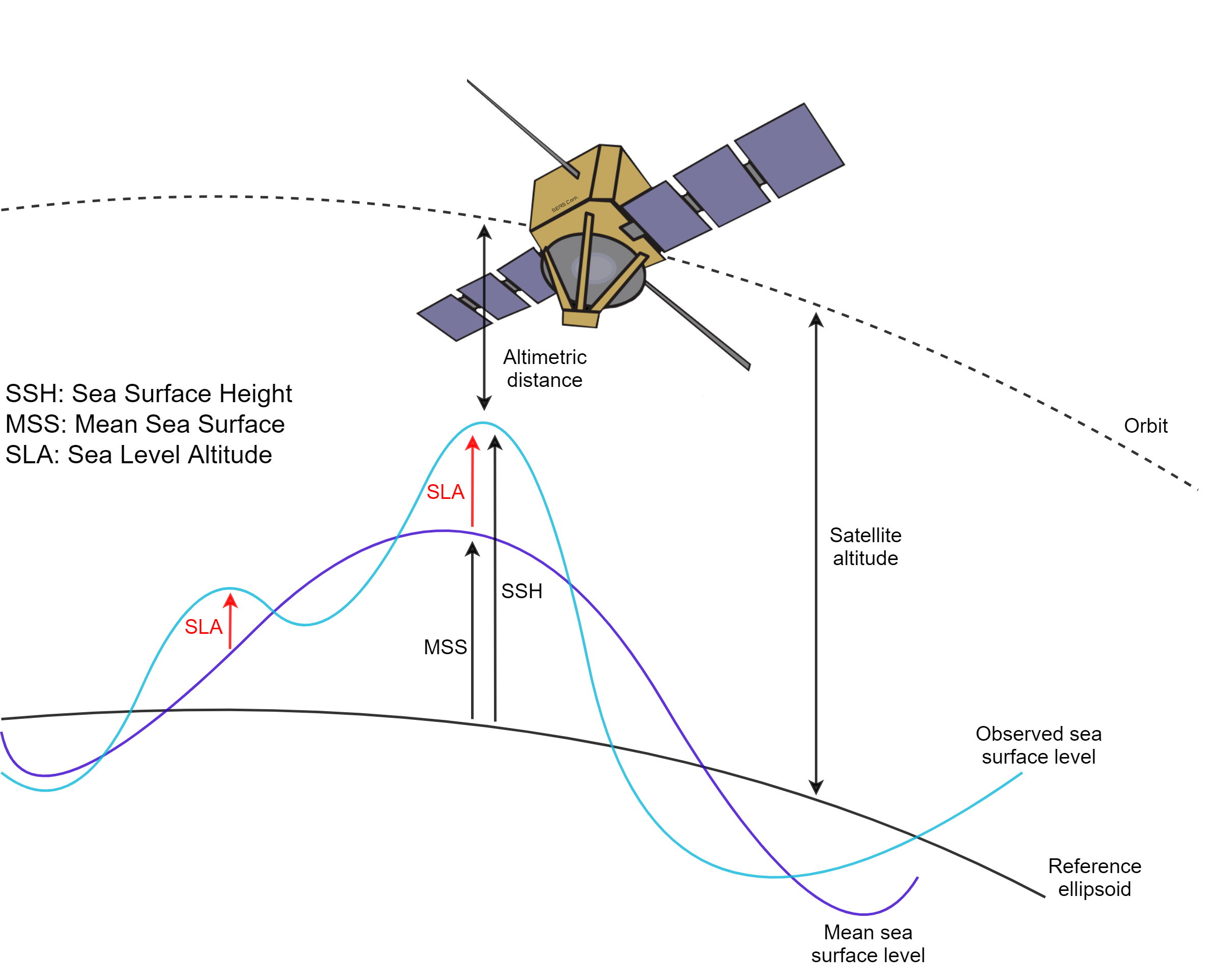}
            \caption[Data collection process]{Data collection process. The image is an author's replica of Figure 1 in \href{https://datastore.copernicus-climate.eu/documents/satellite-sea-level/D3.SL.1-v1.2_PUGS_of_v1DT2018_SeaLevel_products_v2.4.pdf}{Copernicus' Product User Guide and Specification v2.4}.}
        \label{fig:Satellite}
        \end{subfigure}%
        \begin{subfigure}{.5\textwidth}
            \centering
            \includegraphics[width=.9\textwidth]{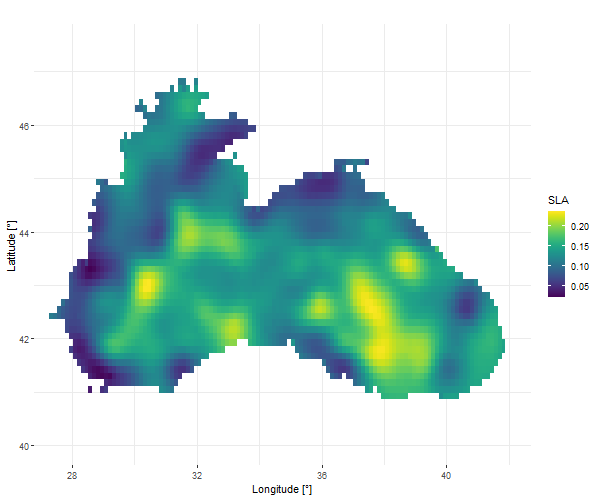}
            \caption[Sea Level Anomaly on 01/01/2018]{Raster representation of the Sea Level Anomaly (in meters) on 01/01/2018}
            \label{fig:BS_plot}
        \end{subfigure}
    \caption{Copernicus Black Sea dataset.}
    \end{figure}

    In this section, we aim to illustrate the application potential of the proposed methodology on a proper case study. 
    %We consider data from Copernicus, the European Union's Earth observation program, which collects vast amounts of global data from satellites and ground-based, airborne, and seaborne measurement systems.
    We analyze a data set from Copernicus Climate Change Service (\href{https://climate.copernicus.eu/}{C3S}), a project operated by the European Center for Medium-Range Weather Forecasts (\href{https://www.ecmwf.int/}{ECMWF}), collecting daily sea level anomalies of the Black Sea in the last twenty years (\citealt{BS_data}).
    Sea level anomalies are measured as the height of water over the mean sea surface in a given time and region. 
    %Anomalies are computed with respect to a twenty-year mean reference period (1993-2012).
    %Satellite altimeters are used to estimate the sea level anomalies with a mapping algorithm dedicated to the Black Sea region. 
    Specifically, altimetry instruments give access to the sea surface height (SSH) above the reference ellipsoid, which is calculated as the difference between the orbital altitude of the satellite and the measured altimetric distance of the satellite from the sea (see \autoref{fig:Satellite}). 
    Starting from this information, Sea Level Anomaly (SLA) is defined as the anomaly of the signal around the Mean Sea Surface component (MSS), which is computed with respect to a 20-year reference period (1993-2012).
    Observations are collected on a spatial raster, with a resolution of $0.125^{\circ}$  both on the longitude and on the latitude axis. 
    Since observations are collected on a geoid, the domain lies on a two-dimensional manifold, however, because both longitude and latitude ranges are very small ($14^{\circ}$ and $7^{\circ}$ respectively), we assume data to be observed on a bidimensional grid.
    The resulting lattice can hence be considered as the Cartesian product of a grid on the longitude axis made by $N_1 = 120$ points and a latitude grid of $N_2 = 56$ points. We refer to $(u_i,v_j)$, with $i=1,\dots,N_1$ and $j=1,\dots,N_2$ as the $(i,j)$-th point of such two-dimensional mesh. 
    %Since the Black Sea does not have a rectangular shape, we will consider each surface to be identically equal to zero outside the perimeter of the sea. 
    Since the Black Sea does not have a rectangular shape, we model data as realization of random surfaces defined on the rectangle circumscribed to the perimeter of the sea, but identically equal to zero outside of it.
    As a consequence, being $\mathcal{B}$ the set of points internal to the perimeter of the Black Sea, we slightly redefine the non conformity measure \ref{eqn:nonconformitymeasure} to become:
    \begin{equation}
    \label{eqn:nonconformitymeasure_application}
        \mathcal{A}(\{z_h: h \in \mathcal{I}_1\}, z) = \esssup_{(u,v) \in [c,d] \times [e,f]} \mathcal{R}(u,v),
    \end{equation}
    %where $\mathcal{R}(u,v)$ is defined as, being $\mathcal{B}$ the set of points belonging to the Black Sea
    where $\mathcal{R}(u,v)$ is defined as:
    \begin{equation}
    \mathcal{R}(u,v) 
        \begin{cases}
            \frac{| y(u,v) - g_{\mathcal{I}_1}(u,v;x_{T+1}) |}{s_{\mathcal{I}_1}(u,v)}, &\text{if } (u,q) \in \mathcal{B},\\
            0, &\text{otherwise}.
        \end{cases}
    \end{equation}
    
    % Since we are settling the study in a Functional Data Analysis framework, we consider the time series $\{SLA_t\}$, without making explicit the dependence on the bivariate domain.
    Hereafter, we will consider the time series $\{SLA_t\}$, without making explicit the dependence on the bivariate domain.
    
    % \begin{figure}
    %     \centering \includegraphics[width=0.5\linewidth]{Pics/Satellite.png}
    %     \caption[Data measurement process]{Data measurement process. The image is an author's replica of Figure 1 in \href{https://datastore.copernicus-climate.eu/documents/satellite-sea-level/D3.SL.1-v1.2_PUGS_of_v1DT2018_SeaLevel_products_v2.4.pdf}{Copernicus' Product User Guide and Specification v2.4}.}
    %     \label{fig:Satellite}
    % \end{figure}
    
    % \begin{figure}
    %     \centering
    %      \includegraphics[width=0.5\textwidth]{Pics/BS_raster.png}
    %     \caption[Sea Level Anomaly on 01/01/2018]{Raster representation of the Sea Level Anomaly (in meters) on 01/01/2018}
    %     \label{fig:BS_plot}
    % \end{figure}

    %%%%%%%%%%%%%%%%%%%%%%%%%%%%%%%%%%%%%%%%%%%%
    %\subsection{Preliminary Analysis} %Preprocessing / Data exploration
    
    \subsection{Data Preprocessing}
    
    %Before applying the proposed forecasting framework, we first preprocess the data in order to work with a 

    If possible, one would preferably forecast directly the time series of Sea Level Anomalies ($SLA_t$). 
    However, in order to estimate the FAR(1) process, we would also like to guarantee stationarity of the time series at our disposal,
    % Since we are embedding our analysis in a functional context, and we aim to estimate a FAR(1) process, we would like to guarantee stationarity of the time series at our disposal.
    % However, 
    and given the particular nature of the dataset, we expect observations to exhibit a strong seasonality as well as a linear trend. Indeed, both tide gauge and altimetry observations show that sea level trends in the Black Sea vary over time (\citealt{black_sea_changes}, \citealt{black_sea_linear_trend}). \citealt{Tsimplis} estimated a rise in the mean sea level of $2.2$ mm/year from 1960 to the early 1990s, while long-track altimetry data indicate that sea level rose at a rate of $13.4 \pm 0.11$ mm/year over 1993–2008 (\citealt{black_sea_Ginzburg}).
    
    %If possible, one would preferably forecast directly the time series of Sea Level Anomalies ($SLA_t$). However, given the nature of the dataset, we expect anomalies to exhibit a strong seasonality as well as a linear trend (\citealt{black_sea_linear_trend}).
    %The Black Sea level has increased by 20 cm in the last 100 years [29].
    %Moreover, both tide gauge and altimetry observations show that sea level trends in the Black Sea vary over time (\citealt{black_sea_changes}). 
    %\citealt{Tsimplis} estimated a rise in the mean sea level of $2.2$ mm/year from 1960 to the early 1990s, while long-track altimetry data indicate that sea level rose at a rate of $13.4 \pm 0.11$ mm/year over 1993–2008 (\citealt{black_sea_Ginzburg}).
    
    %We stress the fact that stationarity is indeed not necessary to obtain valid CP bands, but as proved by \citealt{chernozhukov2018exact}, it is a sufficient condition to guarantee the first assumption of Theorem \autoref{my_theorem}, that we would hence like to be satisfied.

\begin{figure}[]
        \centering
        \begin{subfigure}{.33\textwidth}
            \centering \includegraphics[width=1\linewidth]{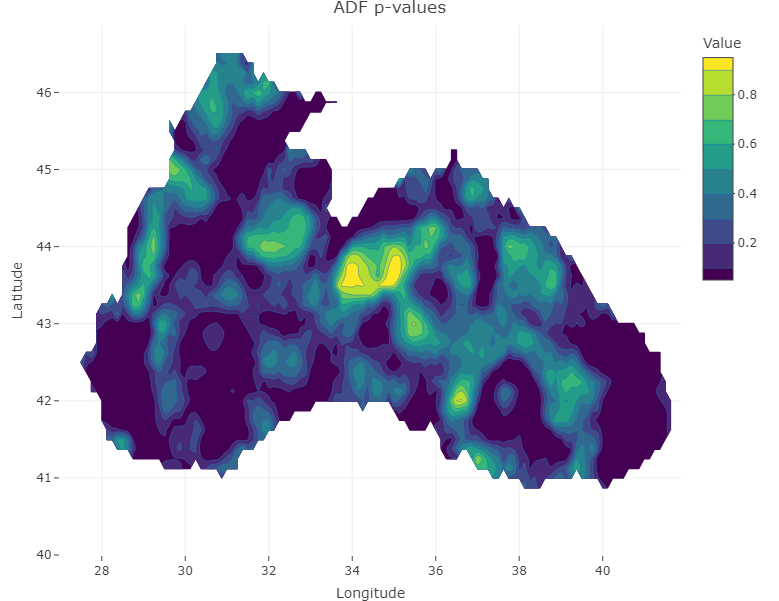}
            \caption{ADF test on $\{SLA_t\}_t$}
            \label{fig:simulation_1_coverage}
        \end{subfigure}%
        \begin{subfigure}{.33\textwidth}
            \centering \includegraphics[width=1\linewidth]{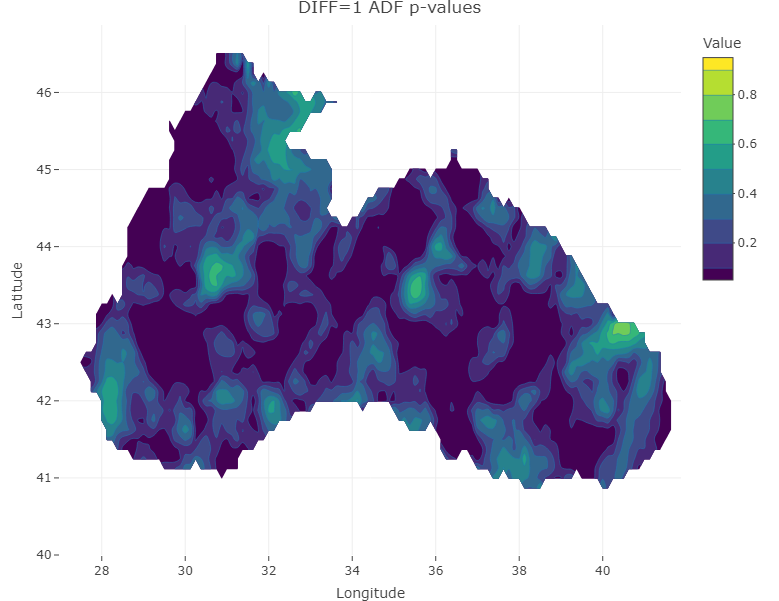}
            \caption{ADF test on $\{\Delta SLA_t\}_t$}
            \label{fig:simulation_1_width}
        \end{subfigure}%
        \begin{subfigure}{.33\textwidth}
            \centering \includegraphics[width=1\linewidth]{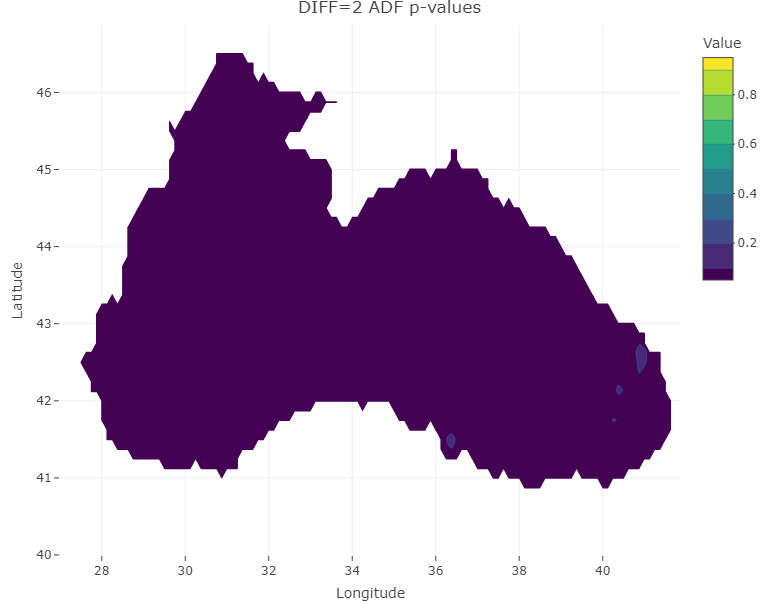}
            \caption{ADF test on $\{\Delta^2 SLA_t\}_t$}
            \label{fig:simulation_2_width}
        \end{subfigure}
        \caption{ADF test's p-values on individual time series. From the left to the right: test on original data, test after one differentiation and test after two differentiations.}
        \label{fig:adf}
    \end{figure}

    In order to further investigate this issue, we should proceed by testing the functional time series $\{SLA_t\}_t$ for stationarity. 
    However, while for one-dimensional Functional Time Series one could resort to the tests proposed by \citealt{FTS_stationarity} or \citealt{Aue_fts_test}, to the best of our knowledge no stationarity test for two-dimensional functional time series has been developed. % and implemented 
    For such reason, and aware of the limits of this approach, we resort to analyzing stationarity of univariate time series $SLA_t(u_i,v_j)$, where each $(u_i,v_j)$ represents a grid point in the lattice.
    We stress the fact that stationarity of individual time series does not guarantee stationarity of the underlying functional process, and this constitutes only a \textit{necessary} and not sufficient condition. 
    Therefore, the goal is not to derive inferential results on the stationarity of the process, but rather to describe the evolution of the process by means of its individual component, and to potentially obtain a better time series to work with.
    We test each univariate time series for stationarity, using the Augmented Dickey Fuller (ADF) test. 
    We report in \autoref{fig:adf} a grid map of the p-values, for $\{SLA_t\}_t$, $\{\Delta SLA_t\}_t$ and $\{\Delta^2 SLA_t\}_t$ , where $\Delta$ and $\Delta^2$ denote respectively one and two differentiations. Despite the fact that we can't make inferential conclusions on the stationarity of the process based on the individual tests, we can see that the original time series exhibit a very non-stationary behavior, and after one differentiation there are many non-stationary locations. After two differentiations, all the individual time series can be confidently considered stationary.
    % The same conclusions have been reached by analyzing individual partial autocorrelation functions and by resorting to a seasonal differentiation.
    We argue that the Functional Time Series might exhibit a behavior similar to one described in terms of stationarity, and hence proceed by differentiating twice the process, defining
    \begin{equation}
        Y_t := \Delta^2 SLA_t = (SLA_{t}-SLA_{t-1}) - (SLA_{t-1}-SLA_{t-2}).
    \end{equation}
    
    We forecast the differentiated time series $Y_t$, and obtain prediction bands for it, using the methodology presented in Section \ref{section:CP}.
    However, in order to provide a better insight into the prediction problem, we need to retrieve results pertaining to the original time series $SLA_t$.
    Specifically, we apply the conformal machinery to the differentiated time series, calling $\hat{Y}_{T+1}$ the forecasted function, and obtaining the prediction band for $Y_{T+1}$:
    \begin{gather*}
        \mathcal{C}_{T,1-\alpha} := \left\{ y \in \mathbb{H}: y(u,v) \in \left[ \hat{Y}_{T+1}(u,v) \pm k^s s_{\mathcal{I}_1}(u,v) \right] \forall (u,v)\right\}.
    \end{gather*}
    Exploiting the fact that:
    \begin{gather*}
        Y_{T+1}(u,v) \in \left[ \hat{Y}_{T+1}(u,v) \pm k^s s_{\mathcal{I}_1}(u,v) \right] \quad \forall (u,v) \\
        \iff \\
        SLA_{T+1}(u,v) \in \left[ \hat{Y}_{T+1}(u,v)+ 2 SLA_T(u,v) - SLA_{T-1}(u,v) \pm k^s s_{\mathcal{I}_1}(u,v) \right] \quad \forall (u,v),
    \end{gather*}
    we define the prediction band for $SLA_{T+1}$ as:
    \begin{gather*}
        \tilde{\mathcal{C}}_{T,1-\alpha} := \left.\{ y \in \mathbb{H}: y(u,v) \in 
        \left[ \hat{Y}_{T+1}(u,v)+ 2 SLA_T(u,v) - SLA_{T-1}(u,v) \pm k^s s_{\mathcal{I}_1}(u,v) \right] \forall (u,v)\right.\}.
    \end{gather*}
    Finally, notice once again that testing the hypothesis underlying the CP procedure is not possible in this setting. We are nevertheless interested in applying the CP scheme together with the forecasting techniques in order to verify their empirical performances.

    \subsection{Study Design}
\begin{figure}
        \centering
        \includegraphics[width=0.4\linewidth]{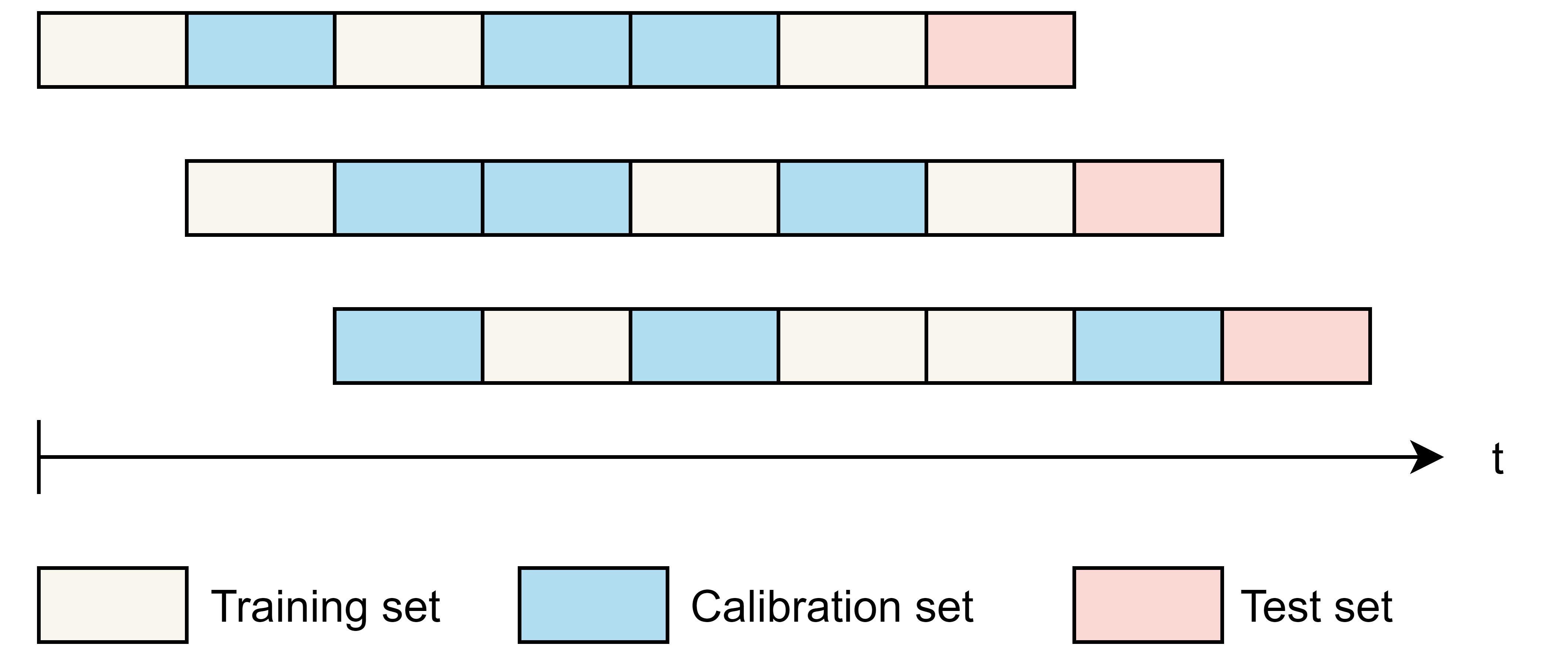}
        \caption[Training-calibration-test split in a rolling window scenario]{Example of the training-calibration-test split in a rolling window scenario.}
        \label{fig:rolling_bs}
    \end{figure}

    The case study employs a rolling estimation framework which recalculates the model parameters on a daily basis and consequently shifts and recomputes the entire training, calibration and test windows by 24 hours, as shown in \autoref{fig:rolling_bs}. 
    As before, we use a random split of data in the training and calibration sets, with split proportion equal to 50\%.
    The significance level $\alpha$ is once again fixed equal to 0.1.
    For each of the 1000 days we aim to predict, we build the corresponding prediction band based on the information provided by the last 99 days, thereby fixing $T=99$. 
    Choosing this sample size provides accurate forecasts and thus small prediction bands, while maintaining reasonable computational times.
    %The grid of points is indeed very dense, therefore a too small sample size may not be suitable to model such complex data.
    %For what concerns the sample size $T$, we will act similarly to Section \ref{subsec:sample_size}, repeating the procedure for increasing sample size, in order to discuss the asymptotic properties of the proposed prediction bands in this more complex scenario. Specifically, $T$ will assume values $19,49,99,499$.
    The size of the blocking scheme is fixed to 1, since, as motivated in Section \ref{subsec:block_size}, this choice produces the narrowest prediction bands, preserving at the same time satisfactory performance in terms of empirical coverage.
    The rolling window is shifted 1000 times, thus iterating for almost three years the forecasting scheme. 
    More specifically, and to allow for reproducibility of subsequent results, we consider a rolling window ranging from 01/01/2017 to 04/01/2020.
    
    The point predictors used throughout this application are those described in Section \ref{section:point_prediction}. 
    The number of Functional Principal Components is set equal to 8.
    % The number of Functional Principal Components is set equal to 8 through a dedicated validation procedure. Given the vast number of observations at our disposal, we can indeed reserve a fraction of them as a validation set, and use it to estimate the optimal number of principal components by means of the Mean Squared Error \eqref{eqn:MSE} between the predicted curve and the actual forthcoming one.
    %By choosing beforehand the number of harmonics rather than selecting them each time by means of the cumulative proportion of variance (as instead done in \ref{section:point_prediction}), we expect to obtain better results in terms of predictive performances.
    % Specifically, historical data from 01/01/2014 to 04/01/2018 are used in order to determine the optimal number of harmonics.?????
    For each shift of the rolling window and for each forecasting algorithm, we check if $SLA_{T+1}$ belongs to $\tilde{\mathcal{C}}_{T,1-\alpha}(x_{T+1})$, and save the size of the corresponding prediction band.
    After collection of results, we calculate the average coverage, and use it to compare performances of the different point predictors in this scenario.
    % We stress the fact that such quantity does not provide a good estimate of the empirical coverage of employed methods, since it is computed from correlated data, and should rather be interpreted as the fraction of time that our method produces a prediction band for the function at $T+1$ containing the actual observed surface $y_{T+1}$. Indeed, by shifting the rolling window by one day at the time, we are inevitably including in the new window all the previous data except one. Nevertheless, a similar setting is often used in practical application, and it is still interesting to compare performances of the different point predictors in this scenario.
    
    %-----------------------------------------------------------
    
    \subsection{Results}
    \label{subsec:BS_results}

\begin{figure}[] 
      \label{ fig7} 
        \begin{subfigure}{.25\textwidth}
            \centering
            \includegraphics[width=\linewidth]{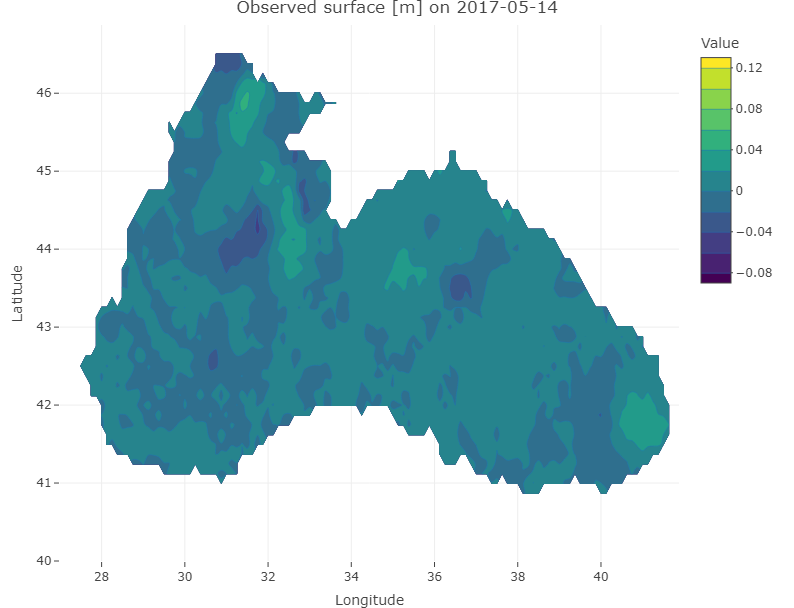} 
            \caption{Observed Surface} 
        \end{subfigure}%%
        \begin{subfigure}{.25\textwidth}
            \centering
            \includegraphics[width=\linewidth]{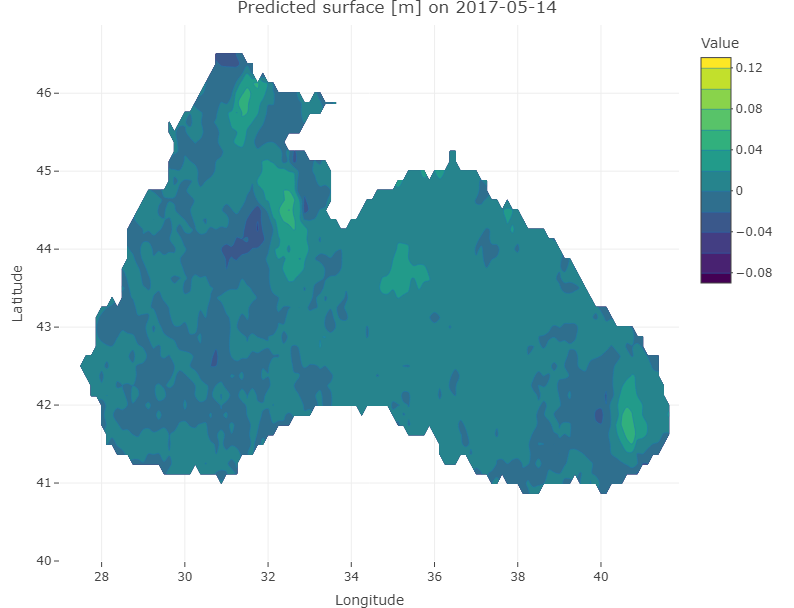} 
            \caption{Predicted Surface} 
        \end{subfigure}%%
        \begin{subfigure}{.25\textwidth}
            \centering
            \includegraphics[width=\linewidth]{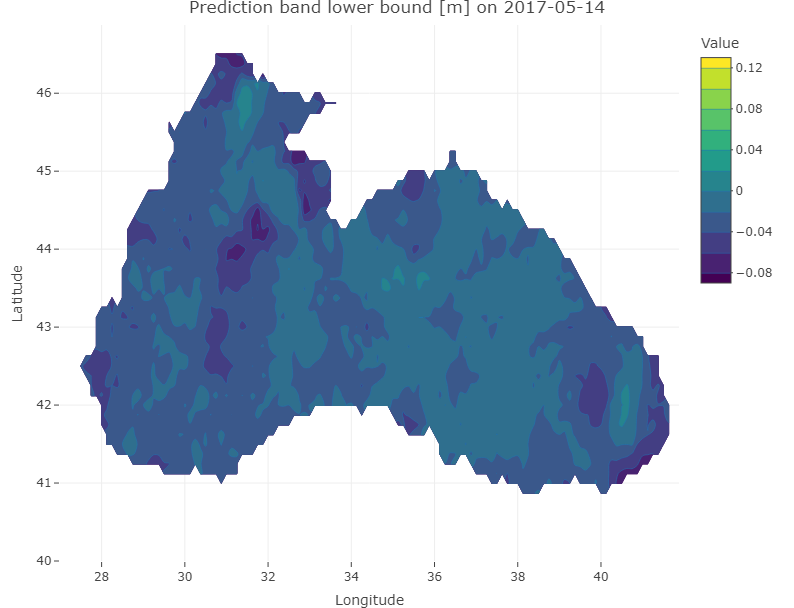} 
            \caption{Prediction Band Lower Bound} 
        \end{subfigure}%%
        \begin{subfigure}{.25\textwidth}
            \centering
            \includegraphics[width=\linewidth]{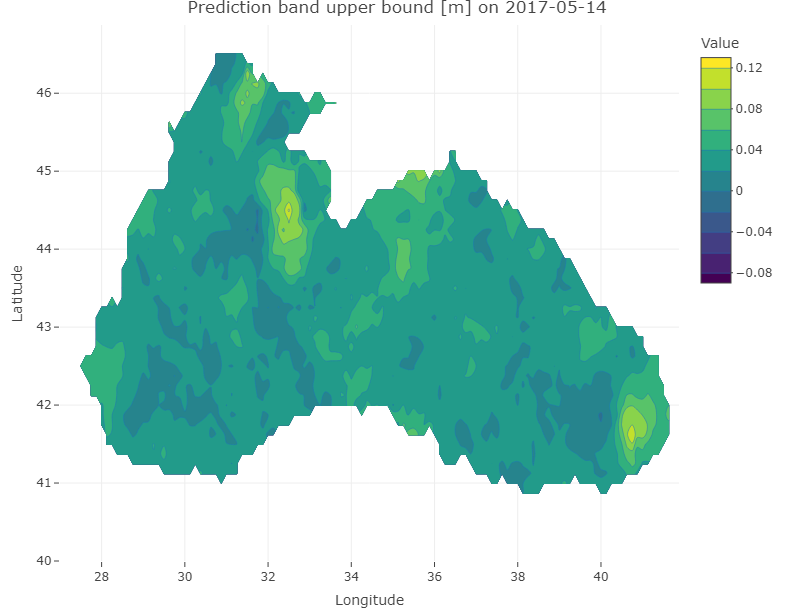} 
            \caption{Prediction Band Upper Bound} 
        \end{subfigure}%%
      \caption{Observed surface, predicted surface, prediction band lower bound, prediction band upper bound on 14/05/2017. Results are obtained using a sample size $T$ equal to 99, $b=1$, $l=48$, $\alpha=0.1$. The employed forecasting algorithm $g_{\mathcal{I}_1}$ is FAR(1)-EK \eqref{eq:EK_estimator}.}
      \label{fig:panel}
    \end{figure}

\begin{figure}[]
        \centering
        \begin{subfigure}{.5\textwidth}
            \centering  \includegraphics[width=1\linewidth]{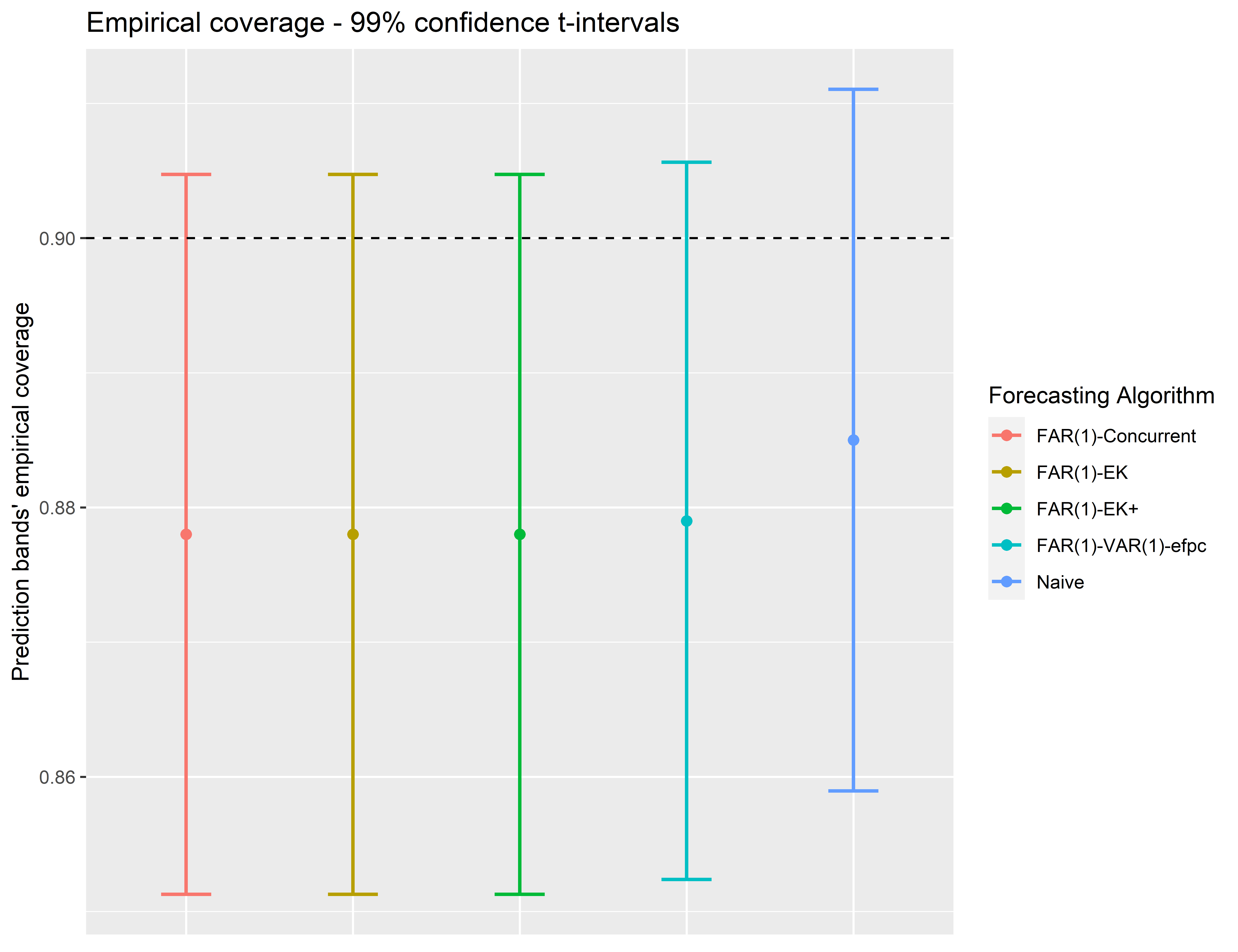}
            \caption{Empirical coverage of CP bands. Dashed line represents nominal coverage $1-\alpha$.}
            \label{fig:BS_coverage}
        \end{subfigure}%
        \begin{subfigure}{.5\textwidth}
            \centering  \includegraphics[width=1\linewidth]{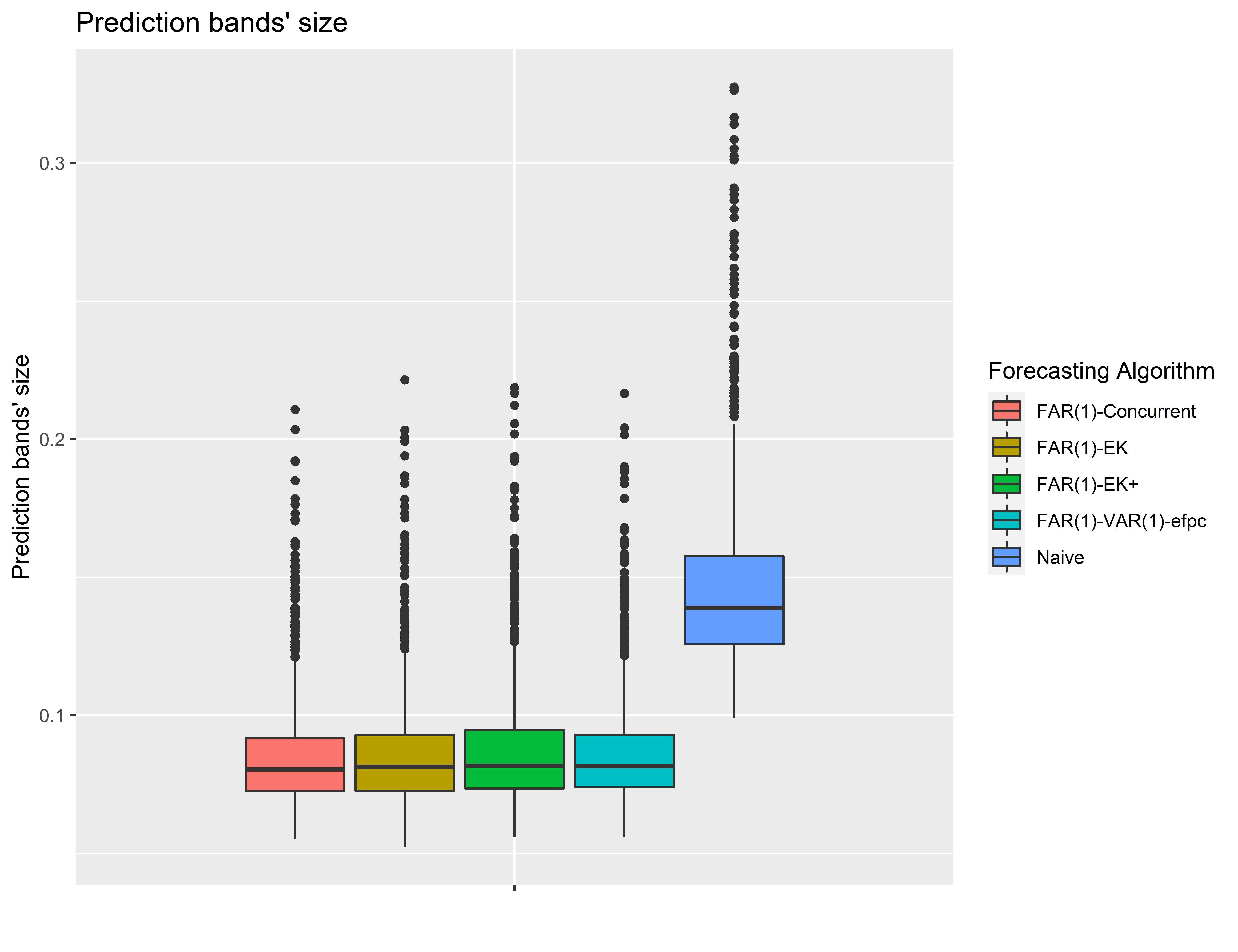}
            \caption{Size of CP bands.}
            \label{fig:BS_width}
        \end{subfigure}
        \caption{Results of the forecasting procedure in a rolling window setting.
        All the methods produce prediction bands with average coverages close to the nominal one.
        For what concerns predictive efficiency, the naive predictor produces the widest bands, whereas the other methods produce instead more efficient bands, with sizes similar to each other.}
        \label{fig:BS_results}
    \end{figure}

    \autoref{fig:panel} outlines
    %In order to visualize results and show the potential applications of our method,
    % we outline in \autoref{fig:panel} 
    observed and forecasted surfaces obtained for one of the day in the rolling window, as long as the lower and upper bounds defining the prediction band.
    For the sake of simplicity, we display only results obtained using the FAR(1)-EK estimator \eqref{eq:EK_estimator}, since, as discussed below, it provides on average the narrowest prediction bands. 
    In order to allow for a more insightful analysis, and to further investigate the evolution of the surfaces, we implemented a dedicated  \href{https://niccolo-ajroldi.shinyapps.io/Black-Sea-Forecasting/}{Shiny App} available online where results can be explored. 
    We report in \autoref{fig:BS_coverage} the average coverage of CP bands obtained across 1000 predictions.
    As in Section \ref{section:4}, we pair such quantity with a 99\% confidence interval. 
    Notice that in this case the confidence interval may be biased, due to the inevitable correlation between data used to construct it, however, we include it in order to assess the dispersion of the average coverage around the mean. 
    Coherently with the results of the simulation study, we can appreciate that in all cases the prediction bands capture the observed surface $y_{T+1}$ approximately $(1-\alpha)\%$ of the times, regardless of the forecasting algorithm used.
    
    For what concerns the size of the prediction bands, the Naive predictor produces by far the widest ones (see \autoref{fig:BS_width}), and, this fact does not reflect in a greater coverage compared to the other methods. 
    On the other hand, prediction bands obtained with autoregressive forecasting algorithms provide narrower prediction regions. Among these, we can see that the non-concurrent FAR(1) is the most performing one, regardless of the way in which it is estimated (namely with FAR(1)-EK, FAR(1)-EK+ or FAR(1)-VAR). Nevertheless, also the concurrent FAR(1) model provides very tight prediction bands, almost comparable with the ones produced by the non-concurrent prediction algorithm.
    
\begin{figure}[] 
        \centering
        \includegraphics[width=.45\linewidth]{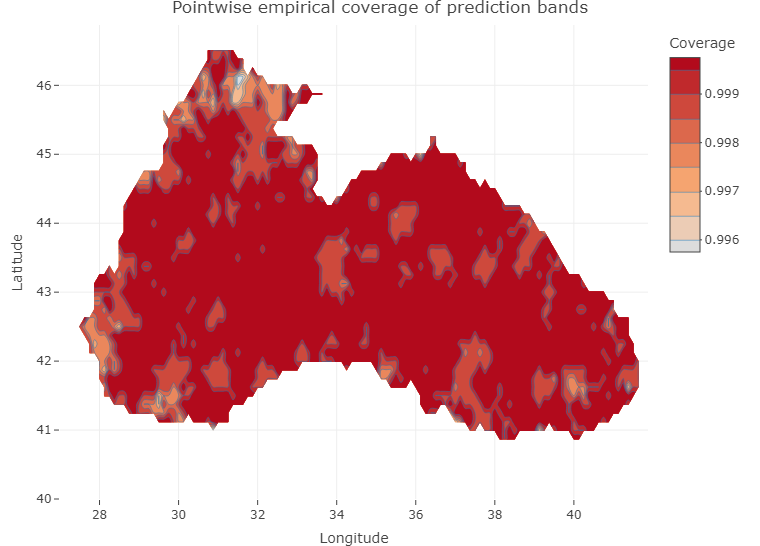}
        \caption{Pointwise coverage of prediction bands across the domain, obtained by counting the number of times that each point falls between prediction bands, and dividing by the total number of time steps in the rolling window framework.} 
        \label{fig:map_cov}
    \end{figure}

\begin{figure}[]
        \centering
        \begin{subfigure}{.5\textwidth}
            \centering  \includegraphics[width=.9\linewidth]{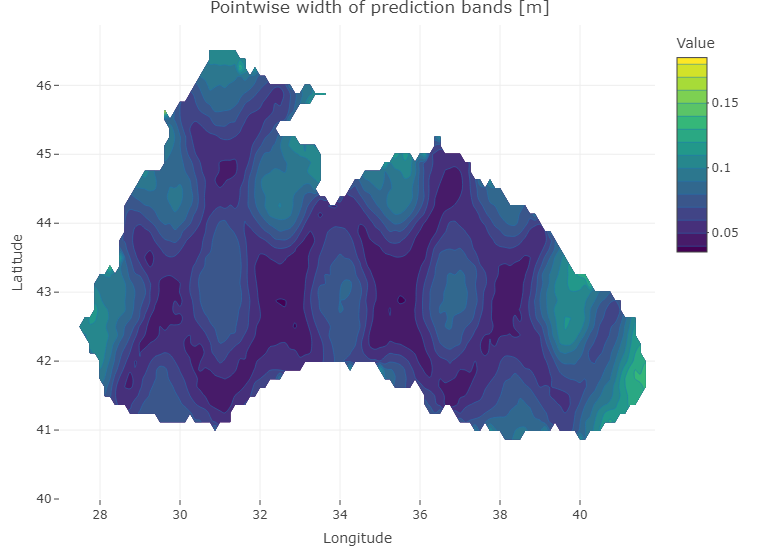}
            \caption{Pointwise average width of CP bands.}
            \label{fig:map_width}
        \end{subfigure}%
        \begin{subfigure}{.5\textwidth}
            \centering  \includegraphics[width=.9\linewidth]{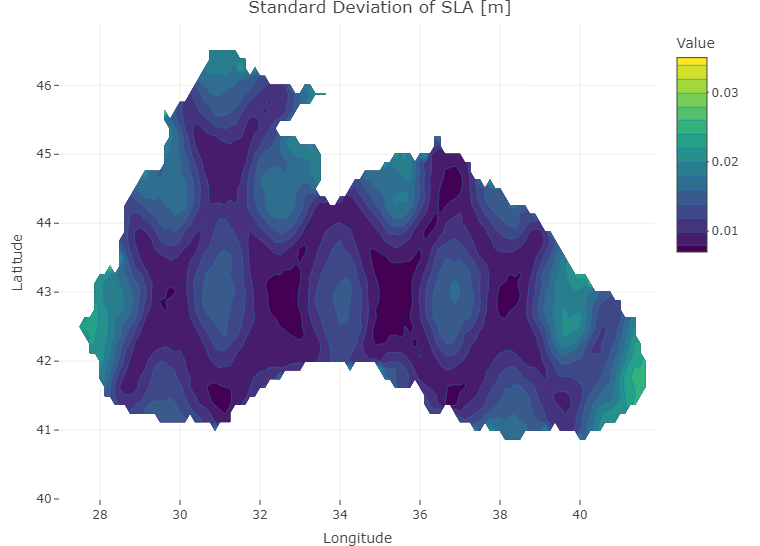}
            \caption{Functional standard deviation of original data.}
            \label{fig:map_std}
        \end{subfigure}
        \caption{Average CP bands' pointwise width and functional standard deviation of original data. The figure on the right denotes a peculiar pattern in the data collection process, which is retrieved in the left panel, due to the choice of using the standard deviation as a modulation function in the CP framework.}
    \end{figure}

    We are also interested in analyzing the \textit{pointwise} properties of CP bands in this scenario. Therefore, we display in \autoref{fig:map_cov} a map of the pointwise coverage of the prediction bands, obtained using FAR(1)-EK. We can appreciate, as expected, a high empirical coverage across the entire domain, emphasizing once again the peculiarity of our approach, which guarantees \textit{global} coverage of the prediction surfaces, reflected by an obvious pointwise coverage higher than the nominal one. We report in \autoref{fig:map_cov} the average width of CP bands, which denotes a peculiar pattern, likely caused by data collection routines. Indeed, we can see from \autoref{fig:map_std}, that a similar behaviour observed in the map of pointwise width can be found by plotting the standard deviation of original data.
    This is coherent with the employed CP framework, since the size of prediction bands depends on the amplitude of the functional standard deviation.

    In conclusion, this  case study confirms the validity of our procedure and proves how a FAR(1) model significantly improves the predictive efficiency even in this more complex scenario.
    
    % \begin{figure}[] 
    %   \label{ fig7} 
    %   \begin{minipage}[b]{0.5\linewidth}
    %     \centering
    %     \includegraphics[width=\linewidth]{Pics/observed.png} 
    %     \caption{Observed Surface} 
    %     \vspace{4ex}
    %   \end{minipage}%%
    %   \begin{minipage}[b]{0.5\linewidth}
    %     \centering
    %     \includegraphics[width=\linewidth]{Pics/predicted.png}
    %     \caption{Predicted Surface} 
    %     \vspace{4ex}
    %   \end{minipage} 
    %   \begin{minipage}[b]{0.5\linewidth}
    %     \centering
    %     \includegraphics[width=\linewidth]{Pics/lower.png} 
    %     \caption{Prediction Band Lower Bound} 
    %     \vspace{4ex}
    %   \end{minipage}%% 
    %   \begin{minipage}[b]{0.5\linewidth}
    %     \centering
    %     \includegraphics[width=\linewidth]{Pics/upper.png} 
    %     \caption{Prediction Band Upper Bound} 
    %     \vspace{4ex}
    %   \end{minipage}
    %   \caption{Observed surface, predicted surface, prediction band lower bound, prediction band upper bound on 14/05/2017. Results are obtained using a sample size $T$ equal to 99, $b=1$, $l=48$, $\alpha=0.1$. The employed forecasting algorithm $g_{\mathcal{I}_1}$ is FAR(1)-EK \eqref{eq:EK_estimator}.}
    %   \label{fig:panel}
    % \end{figure}

\section{Conclusions and Further Developments}
\label{section:conclusions}

In this work, we introduce a mathematical framework for probabilistic forecasting of two-dimensional Functional Time Series. 
Leveraging the CP scheme developed by \citealt{chernozhukov2018exact} and \citealt{diquigiovanni2021_FTS} and adapting it to the 2D-FTS setting, we propose technique for quantifying uncertainty when predicting time evolving surfaces.
In order to provide point prediction of surfaces, we model functions through a Functional autoregressive process of order one, extending the mathematical theory of autoregressive processes to allow for bivariate functions. Estimations techniques for the FAR(1) are presented and compared 
We test the benefits and limits of the proposed approach, first on synthetic data and then on a real novel time series dataset, collecting daily observations of sea level anomaly over the Black Sea. 
Empirical results proved the validity of the methodology on non-synthetic data.
We acknowledge that in applying the proposed procedure to the case study, we had to introduce some simplifications due to the novelty of the subject and the limited amount of work on two-dimensional Functional Time Series. We hope that this work will encourage the development of novel analysis techniques for 2D functional data, such as stationarity tests and other forecasting tools.
Finally, throughout the work, we limited the analysis to the FAR(p), with $p=1$, as motivated in \autoref{section:point_prediction}. 
One may consider a FAR(p) model, with $p>1$. Whereas it is not straightforward to extend the estimation of $\Psi_M$ \eqref{eq:EK_estimator} to a FAR(p) model with $p>1$,  both the concurrent estimator \eqref{eqn:FAR_concurrent} and the estimator based on an expansion of FPC's, may be easily adapted to the case $p>1$. The problem can in fact be rendered as a standard functional linear regression and solved by using the many off-the-shelf methods apt to the task and present in the literature \citep[see e.g.][]{chiou2016multivariate}. Finally, notice that thanks to the flexibility of our approach, a modification of such kind can be easily achieved by changing the point predictor $g_{\mathcal{I}_1}$ only, while preserving the desirable properties of prediction bands.

\section*{Acknowledgments}
The present research has been partially supported by MUR, grant Dipartimento di Eccellenza 2023-2027 and by Accordo Attuativo ASI-POLIMI ``Attività di Ricerca e Innovazione" n. 2018-5-HH.0, collaboration agreement between the Italian Space Agency and Politecnico di Milano.
M.F. acknowledges the support of the JRC Centre of Advanced Studies CSS4P - ``Computational Social Science for Policy".
% The authors would also like to thank Edoardo Marchionni for the fruitful discussions on basis functions.

%%%%%%%%%%%%%%%%%%%%%%%%%%%%%%%%%%%%%%%%%%%%%%%%%%%%%%%%%%%%%%

%\section*{Code}
%All the analysis are implemented using the R Programming Language %(\citealt{R}). \\
%Codes from \citealt{diquigiovanni2021importance}, implementing Conformal %Inference in the functional setting, have been adapted to allow for %functions defined on a bivariate domain.
%Functional Principal Component Analysis in the two-dimensional %functional setting, estimation methods for the Functional Autoregressive %Process of order one and the other forecasting algorithms are all %implemented from scratch, as long as simulations and tests. \\
%Codes are so far not publicly available, but the authors are at disposal %for any clarification on the implementation details.

%Simulations are performed in a parallel fashion employing the package \texttt{snowfall}.

%%%%%%%%%%%%%%%%%%%%%%%%%%%%%%%%%%%%%%%%%%%%%%%%%%%%%%%%%%%%%%

%\newpage

\bibliographystyle{elsarticle-harv} 
\bibliography{bibliography}

%%%%%%%%%%%%%%%%%%%%%%%%%%%%%%%%%%%%%%%%%%%%%%%%%%%%%%%%%%%%%%

%\newpage
\appendix

%%%%%%%%%%%%%%%%%%%%%%%%%%%%%%%%%%%%%%%%%%%%%%%%%%%%%%%%%%%%%%

\section{Split ratio and type of split}
\label{appendix_split}

    The choice of the split ratio is non-trivial and has motivated discussion in the statistical community. Including more data in the training set improves the estimation of the point predictor $\hat{g}_{\mathcal{I}_1}$. %t the same time, having fewer data in the calibration set produces a very rough p-value function, which mught resulti in greater actual coverage with respect to the nominal one. 
    At the same time, having few data points in the calibration set produces a very rough p-value function \eqref{eqn:p_value_function}, resulting in potential greater actual coverage with respect to the nominal one. 
    This trade-off problem is enhanced in the time series context, in which one would like to have both training and calibration sets as large as possible, since asymptotic validity is guaranteed when both $l$ and $m$ go to infinity. 
    Throughout this work, the training-calibration ratio is fixed equal to 50\%-50\%, as commonly suggested in literature. 
    Moreover, we stress the fact that the split is random. This clearly introduce variability in the procedure since results depend on the particular division of data.
    We acknowledge a recent advancement in this direction, called Multi Split Conformal Prediction (\citealt{multisplitCP}), which aggregates single split %conformal prediction 
    CP intervals across multiple splits. 
    
\begin{figure}
        \centering
        \begin{subfigure}{.45\textwidth}
            \centering \includegraphics[width=1\linewidth]{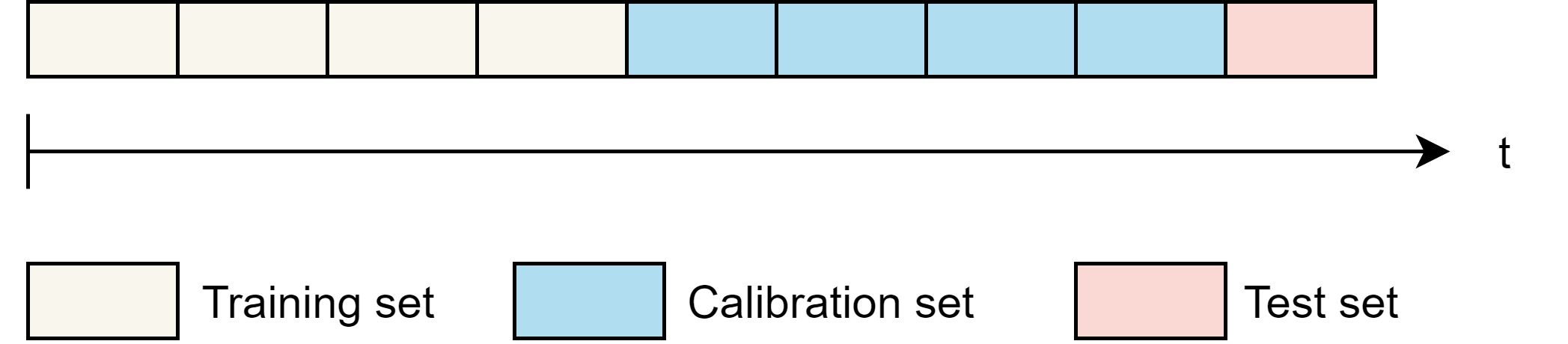}
            \caption{Consecutive split}
            \label{fig:consecutive_split}
        \end{subfigure}%
        \begin{subfigure}{.45\textwidth}
            \centering \includegraphics[width=1\linewidth]{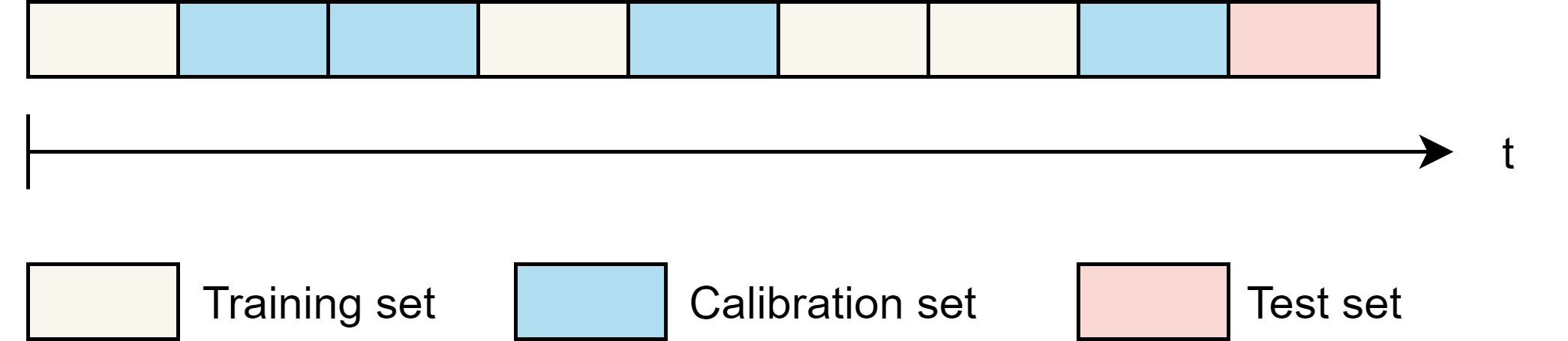}
            \caption{Random split}
            \label{fig:random_split}
        \end{subfigure}
        \caption{Possible types of split with $T=8$, $m=l=4$.}
        \label{fig:test}
    \end{figure}

    Another interesting question regarding the type of split comes up in the time series context. Due to the lack of exchangeability, two different subdivisions are possible in this framework.
    A first choice could consist in a sequential division of data, where the split point is no longer random, but is a result of the training-calibration proportion (see \ref{fig:consecutive_split}). \citealt{Wisniewski2020} applied this scheme  in a rolling window fashion to forecast Market Makers' Net Positions.
    While this choice may seem more consistent in the presence of temporal dependence, since it does not split subsequent observations in two different sets, it may lead to very biased results if the training size $m$ is very small or if data present a different trend or seasonal component in the training and calibration sets. The interested reader may refer to \citealt{Kath} for a more comprehensive discussion on this topic.
    %This conjecture, also discussed by \citealt{Kath}, is investigated in more details in Appendix ?? with an ad-hoc simulation study.
    All in all, in order to make the model more robust, we preferred to split data randomly (\ref{fig:random_split}).

%%%%%%%%%%%%%%%%%%%%%%%%%%%%%%%%%%%%%%%%%%%%%%%%%%%%%%%%%%%%%%

\section{FAR(1) estimation and adaptation to CP}
\label{appendix_EK}
    
    Hereafter, we will assume $\{Y_t\}_{t=1}^{T} \subset \mathcal{L}^4(\Omega, \mathcal{F},\mathbb{P})$ and consider only zero-mean stationary time series.
    The covariance operator $\Gamma_0: \mathbb{H} \rightarrow \mathbb{H}$ and the lag-1 autocovariance operator $\Gamma_1$ can thus be defined as:
    \begin{gather}
        \Gamma_0 x = \mathbb{E}[\langle Y_t, x \rangle Y_t] \qquad almost \; everywhere \; (a.e.), \\
        \Gamma_1 x = \mathbb{E}[\langle Y_t, x\rangle Y_{t+1}] \qquad a.e.
    \end{gather}
    Extending the work of \citealt{horvath2012inference} to a different functional space, we can derive the following estimators for such operators:
    \begin{gather}
        \hat{\Gamma}_0 x = \frac{1}{T} \sum_{t=1}^T \langle Y_t, x \rangle Y_t \qquad a.e., \\
        \hat{\Gamma}_1 x = \frac{1}{T-1} \sum_{t=1}^{T-1} \langle Y_t, x \rangle Y_{t+1} \qquad a.e..
    \end{gather}
    
    Under rather general weak dependence assumptions these estimators are $\sqrt{n}$-consistent. 
    One may, for example, adopt the concept of $\mathcal{L}^p$-$m$-approximability introduced in \citealt{hormann_kokoska} to prove that $\mathbb{E}[||\hat{\Gamma_0}-\Gamma_0||^{2}]= \mathcal{O}(n^{-1})$, where $||.||$ is the operatorial norm: $||F|| := \sup_{||x||_{\mathbb{H}}=1} ||Fx||_{\mathbb{H}}$ for any linear bounded operator $F:\mathbb{H}\rightarrow\mathbb{H}$.
    
    In order to derive estimator \eqref{eq:EK_estimator} of the FAR(1), we first consider the following operatorial equation:
    \begin{equation}
    \label{eqn:Gamma1_ideal}
        \Gamma_1 = \Psi \Gamma_0.
    \end{equation} 
    A natural idea may consist in computing estimators $\hat{\Gamma}_0$, $\hat{\Gamma}_1$ from historical data and defining then $\hat{\Psi} = \hat{\Gamma}_0  \hat{\Gamma}^{-1}_0$.
    Unfortunately, the inverse operator $\Gamma^{-1}_0$ is unbounded on $\mathbb{H}$ (\citealt{horvath2012inference}),
    % potrei anche dimostrarlo da me, perché Horvath lo fa nel caso 1D! In tal caso aggiungo una nota a piè pagina
    however, thanks to $\Gamma_0$ being a symmetric, compact, positive-definite operator, one can exploit its spectral decomposition to introduce a pseudo-inverse operator $\Gamma_{0,M}^{-1}$, defined as: 
    \begin{equation}
    \label{eqn:pseudo-inverse}
        \Gamma_{0,M}^{-1}x = \sum_{j=1}^{M} \lambda_j^{-1} \langle x, \xi_j \rangle \xi_j \qquad a.e.,
    \end{equation}
    where $\xi_1, \dots ,\xi_M$ are the first M normalized functional principal components (FPC's), $\lambda_1, \dots, \lambda_M$ are the corresponding eigenvalues and $\langle x, \xi_1 \rangle, \dots, \langle x, \xi_M \rangle$ are the scores of $x$ along the FPC's.
    We formally define $\xi_i$ and $\lambda_i$ as eigenfunctions and eigenvalues that solve the functional equation:
    \begin{equation}
        \Gamma_0 \xi_i = \lambda_i \xi \qquad i=1,\dots,M.
    \end{equation}
    
    We can now combine \eqref{eqn:Gamma1_ideal} and \eqref{eqn:pseudo-inverse}, plugging in estimated eigenfunctions and eigenvalues and calling $\hat{\Gamma}_{0,M}^{-1}$ the resulting estimator of $\Gamma_{0,M}^{-1}$, to finally derive:
    
    \begin{equation}
        \hat{\Psi}_M = \hat{\Gamma}_{1} \hat{\Gamma}_{0,M}^{-1}x,
    \end{equation}
    \begin{equation}
    \label{eq:EK_estimator_APP}
        \hat{\Psi}_{M} x = \frac{1}{T-1} \sum_{i,j = 1}^{M} \sum_{t=1}^{T} \hat{\lambda}_j \langle x, \hat{\xi}_j \rangle \langle Y_{t-1}, \hat{\xi}_j \rangle \langle Y_{t}, \hat{\xi}_i \rangle \hat{\xi}_i
        \qquad a.e..
    \end{equation}
    Notice that the operator $\hat{\Gamma}_{0,M}^{-1}$ is bounded on $\mathbb{H}$ if $\hat{\lambda}_j$ are strictly greater than zero for $j=1, \dots ,M$. Nevertheless, even if such condition is met, in practice one should cautiously select the number of principal components $M$, because very small eigenvalues will result in very high reciprocals $\hat{\lambda}^{-1}_j$, providing in practice unbounded estimates of $\Gamma_{0,M}^{-1}$.
    Such observation motivated \citealt{didericksen} to add a positive baseline to the estimated eigenvalues $\hat{\lambda}_j$. This small modification improves the estimation of the operator $\Psi$, and most importantly, contributes to weaken the dependency of $\hat{\Psi}_M$ on $M$. 

    We now aim to adapt the previous estimator \eqref{eq:EK_estimator_APP} to the conformal inference setting. The goal is to accommodate the forecasting algorithm in order to estimate the point predictor $g_{\mathcal{I}_1}$ from the training set only. 
    As a preliminary step, given that the FAR(1) model has been presented for mean-centered data, one has to estimate the mean function $\hat{\mu}_{\mathcal{I}_1}$ from the training set only and consequently center all the observations in the training and calibration set around $\hat{\mu}_{\mathcal{I}_1}$.
    If the sample size at disposal is sufficiently large and if the stationarity assumption is fulfilled, it should make no great difference to estimate the population function $\mu$ with the sample mean $\hat{\mu}=\frac{1}{T}\sum_{t=1}^T Y_t$ or with its restriction on the training set $\hat{\mu}_{\mathcal{I}_1}=\frac{1}{m}\sum_{t \in \mathcal{I}_1} Y_t$.
    Another fundamental step is the estimation of functional principal components. Since in the CP framework we are allowed to use only the information from the training set in order to compute $\hat{\xi}_1, \dots, \hat{\xi}_M$, it is then natural to employ only training data $\{z_h: h \in \mathcal{I}_1\}$ in such estimation routine.
    
    In order to obtain estimator $\hat{\Psi}_M$ \eqref{eq:EK_estimator}, one has to compute $\hat{\Gamma}_1$ and $\hat{\Gamma}_{0,M}^{-1}$ from the training set. While the computation of the sample pseudo-inverse of the autocovariance estimator is straightforward:
    \begin{equation}
      \hat{\Gamma}_{0,M}^{-1} x = \frac{1}{m} \sum_{j=1}^{M} \hat{\lambda}_j \langle x, \hat{\xi}_j \rangle \hat{\xi}_j \qquad a.e.,
    \end{equation}
    the CP counterpart of $\hat{\Gamma}_{1}$ is more delicate and requires further discussion.
    Recall indeed that the classical estimator for the lag-1 autocovariance operator from $Y_1, \dots, Y_T$ is:
    \begin{gather}
        \label{eqn:repr1}
        \hat{\Gamma}_1 x = \frac{1}{T-1} \sum_{t =1}^{T-1} \langle Y_t, x \rangle Y_{t+1} = \frac{1}{T-1} \sum_{t =2}^{T} \langle Y_{t-1}, x \rangle Y_t.
    \end{gather}
    In the CP setting, however, we could define three different estimators for $\Gamma_1$:
    \begin{gather}
        \label{eqn:repr1CP}
        \hat{\Gamma}_1 x = \frac{1}{m-1} \sum_{t\in \mathcal{I}_1[1:m-1]} \langle Y_t, x \rangle Y_{t+1}, \\
        \label{eqn:repr2CP}
        \hat{\Gamma}_1 x = \frac{1}{m-1} \sum_{t\in \mathcal{I}_1[2:m]} \langle Y_{t-1}, x \rangle Y_{t}, \\
        \label{eqn:repr3CP}
        \hat{\Gamma}_1 x = \frac{1}{m} \sum_{t\in \mathcal{I}_1} \langle Y_{t-1}, x \rangle Y_{t},
    \end{gather}
    where $x \in \mathbb{H}$ and $\mathcal{I}_1[i:j]$ contains the indices from the $i$-th to the $j$-th element of $\mathcal{I}_1$.
    Notice that the three estimators differ because if $t \in \mathcal{I}_1$, we have no assurance that $\{t-1\} \in \mathcal{I}_1$ or even $\{t+1\} \in \mathcal{I}_1$.
    We stress also the fact that the third operator is well-defined only if we reserve a burn-in set of length 1 at the front of the time series, in such a way that, if $\{2\} \in \mathcal{I}_1$, we can still compute the estimator.
    Among the three options, we prefer the third one \eqref{eqn:repr3CP}, since it averages over a larger set.
    One may also argue that such estimators are not coherent with the CP setting, since they are inevitably based on data from the calibration set. However, as mentioned before, we are considering the time series of \textit{regression pairs} $Z_t = (X_t, Y_t)$, $t=1,\dots,T$ (\citealt{chernozhukov2018exact}). The key consideration is that, according to the FAR(1) model, we use as regressors the lagged version of the time series, namely $X_t=Y_{t-1}$ and the regression couples becomes $Z_t = (Y_{t-1}, Y_t)$, for each $t=1,\dots,T$.
    From this perspective, one could rephrase the definition of the sample covariance operator by making explicit its dependence from the regressor $X_t$ instead of $Y_{t-1}$:
    %introducing the lag operator $L$, finding:
    \begin{align}
        \hat{\Gamma}_1 x &= \frac{1}{m-1} \sum_{t\in \mathcal{I}_1} \langle  Y_{t-1}, x \rangle Y_{t} = \frac{1}{m} \sum_{t\in \mathcal{I}_1} \langle  X_{t}, x \rangle Y_{t}.
    \end{align}
    It is then straightforward to derive the estimator $\hat{\Psi}_{M,\mathcal{I}_1}$:
    \begin{align}
        \hat{\Psi}_{M,\mathcal{I}_1} x &= \frac{1}{m} \sum_{i,j = 1}^{M} \sum_{t \in \mathcal{I}_1} \hat{\lambda}_j \langle x, \hat{\xi}_j \rangle \langle Y_{t-1}, \hat{\xi}_j \langle Y_{t}, \hat{\xi}_i \rangle \hat{\xi}_i = \frac{1}{m} \sum_{i,j = 1}^{M} \sum_{t \in \mathcal{I}_1} \hat{\lambda}_j \langle x, \hat{\xi}_j \rangle \langle X_{t}, \hat{\xi}_j \langle Y_{t}, \hat{\xi}_i \rangle \hat{\xi}_i \qquad a.e..
        \label{eqn:psi_explicit}
    \end{align}
    The point predictor finally becomes:
    \begin{equation}
        \hat{Y}_{T+1} = g_{\mathcal{I}_1}(u,v;X_{T+1}) = (\hat{\Psi}_{M,\mathcal{I}_1} X_{T+1})(u,v) = (\hat{\Psi}_{M,\mathcal{I}_1} Y_{T})(u,v)
    \end{equation}
    %$ \hat{Y}_{T+1} = g_{\mathcal{I}_1}(u,v;X_{T+1}) = (\hat{\Psi}_{M,\mathcal{I}_1} X_{T+1})(u,v) = (\hat{\Psi}_{M,\mathcal{I}_1} Y_{T})(u,v)$.
    
    In order to compute \eqref{eqn:psi_explicit}, it is first mandatory to project the calibration set onto the EPFC's in order to compute the scores $\langle X_{h}, \hat{\xi}_j \rangle $ for $h \in \mathcal{I}_2$. Notice that, in the non-Conformal setting, such step comes for free when performing FPCA. In the CP context, however, EPFC's are computed from the training set only, therefore, in order to access the scores of the calibration set, one needs to explicitly add this projection step.
        
%%%%%%%%%%%%%%%%%%%%%%%%%%%%%%%%%%%%%%%%%%%%%%%%%%%%%%%%%%%%%%%%%%%%%%%%
    
\section{FPCA for Two-Dimensional Functional Data}
\label{appendix_FPCA}
    
    A fundamental aspect in the design of many forecasting algorithms is the estimation of functional principal components (FPC's) $\{\xi_i\}{i \in \mathbb{N}}$. 
    We define FPC's as functions $\xi_i \in \mathcal{L}^2([c,d] \times [e,f])$ solving the functional equation:
    \begin{equation}
        \Gamma_0 \xi = \lambda \xi.
    \end{equation}
    In practice, we can only estimate the first $M\in\mathbb{N}$ eigenfunctions, implicitly performing dimensionality reduction. The choice of $M$ is non-trivial and depends on the application framework.
    Whereas \citealt{kargin} suggested selecting it in a cross-validation setting, \citealt{Aue} proposed a fully automatic criterion for choosing the number of principal components in terms of predictive performances.
    Plugging in the estimator of $\Gamma_0$, we define estimated eigenfunctions and eigenvalues as solutions of:
    \begin{equation}
    \label{eqn:eigenproblem_functional}
        \hat{\Gamma}_0 \hat{\xi} = \hat{\lambda} \hat{\xi}.
    \end{equation}
    On a theoretical point of view, we would like to guarantee that population eigenfunctions can be consistently estimated by empirical eigenfunctions even in the non-iid framework of Functional Time Series.
    We refer to Theorem 16.2 in \citealt{horvath2012inference}, which provides asymptotic arguments for such question.
    
    The following subsections are dedicated to the estimation of eigenfunctions and eigenvalue in the two-dimensional functional case. Extending the work of \citealt{ramsay2005functional}, we present two different estimation procedures, based respectively on a discretization of the functions to a fine grid and on a linear expansion of data on a finite set of basis functions. 
    Among the two alternatives, we would resort to the function discretization. 
    Indeed, such choice does not require the selection of a specific type of basis and not even the number of basis to employ, which are not trivial problem-dependent questions. 
    Moreover, notice that also the discretization procedure can be seen as a particular case of the basis expansion, using as basis system indicator functions on the grid points.
    Furthermore, in the subsequent, \ref{sec:comparison_FPCA} we demonstrate with a simulation study that there is no significant evidence to prefer one method against the other in terms of estimation quality. % and of predictive performances.
    
    We want to stress the fact that our methodology for FPCA is general, it works for two-dimensional functional data regardless of the presence of temporary dependence between observations. Not modeling  the serial dependence structure does not invalidate the PCA procedure, but we still have to require that the dynamic is stationary in order for the covariance estimation to make sense and thus to provide meaningful estimates.
    
    %\subsection{Estimation}
    \subsection{Grid Discretization}
    
        Consider a grid discretization $\{u_i\}_{i=1,\dots,N_1}$ of $[c,d]$ and $\{v_j\}_{j=1,\dots,N_2}$ of $[e,f]$, let $\omega_1 = \frac{1}{N_1}$, $\omega_2 = \frac{1}{N_2}$. For any point $(u_i,v_j)$ of the discretized grid, the lhs of the functional eigenequation \eqref{eqn:eigenproblem_functional} can be rewritten as:
        \begin{align}
            \hat{\Gamma}_0 \hat{\xi} (u_i, v_j) &= \int_c^d \int_e^f \hat{\gamma}_0(u_i,v_j;w,z) \hat{\xi}(w,z) dw dz \approx \\
            &\approx \omega_1 \sum_{l=1}^{N_1} \int_e^f \hat{\gamma}_0(u_i,v_j; u_l,z) \hat{\xi}(u_l,z) dz \approx \\
            \label{eqn:double_sum}
            &\approx \omega_1 \omega_2 \sum_{l=1}^{N_1} \sum_{m=1}^{N_2} \hat{\gamma}_0(u_i,v_j;u_l,v_m) \hat{\xi}(u_l,v_m). %\\
            %&= \omega_1 \omega_2 \bm{\Gamma_0} \bm{\xi}
        \end{align}
        By defining $N:=N_1 N_2$ and introducing a bijection $\zeta: \{1,\dots,N_1\}\times\{1,\dots,N_2\} \rightarrow \{1,\dots,N\}$, we can vectorize the two-dimensional grid. % in such a way that $\theta_{\zeta(i,j)}=(u_i,v_j)$.
        Therefore, we can store observed data into a bidimensional matrix, and proceed with a usual multivariate analysis. Let $\mathbb{Y} \in \mathbb{R}^{T \times N}$ be defined as $\mathbb{Y}[t,\zeta(i,j)]=y_t(u_i,v_j)$.
        We hence introduce the estimated variance-covariance matrix of the just defined multivariate dataset: $\hat{\bm{\Gamma}}_0 \in \mathbb{R}^{N\times N}$, $\hat{\bm{\Gamma}}_0 = \frac{1}{T} \mathbb{Y}^T\mathbb{Y}$. Notice that $\hat{\bm{\Gamma}}_0[\zeta(i,j),\zeta(l,m)]=\hat{\bm{\gamma}}_0(u_i,v_j;u_l;v_m)$. Let also $\hat{\bm{\xi}} \in \mathbb{R}^{N}$, $\bm{\xi}[\zeta(i,j)] = \xi(u_i,v_j)$.
        The eigenequation can thus be rewritten in the following matricial form:
        \begin{equation}
            \omega_1 \omega_2 \hat{\bm{\Gamma}}_0 \hat{\bm{\xi}} = \hat{\lambda} \hat{\bm{\xi}}.
        \end{equation}
        It is then straightforward to find the eigenvalues $\rho$ and eigenvectors $\bm{\theta}$ of the matrix $\bm{\Gamma}_0$ and to derive $\hat{\lambda} = \omega_1 \omega_2 \rho$ and $\hat{\bm{\xi}}=\omega_1^{-1/2} \omega_2^{-1/2} \bm{\theta}$. Finally, to obtain an approximate eigenfunction $\hat{\xi}$ from discrete values $\hat{\bm{\xi}}$, we can use any convenient interpolation method.

    \subsection{Basis Expansion}
    \label{sec:FPCA_basis}

        Let $\{g_i\}_{i \in \mathbb{N}}$ be a basis system for $\mathcal{L}^2([c,d])$ and $\{h_j\}_{j \in \mathbb{N}}$ a basis system for $\mathcal{L}^2([e,f])$.
        Consider now the tensor product basis $\{g_i \otimes h_j\}_{i,j}$, where $g_i \otimes h_j = g_i h_j$. Unfortunately, the space spanned by the tensor product basis is a proper subset of $\mathbb{H}=\mathcal{L}^2([c,d] \times [e,f])$, however, one could also prove that such subspace is \textit{dense} in $\mathbb{H}$, thus arguing that the tensor product basis system is sufficient to model functions of $\mathbb{H}$. 
        Therefore, we assume that each $x \in \mathbb{H}$, admits the decomposition:
        \begin{equation}
        \label{eqn:basis_dec_long}
            x(u,v) = \sum_{i \in \mathbb{N}} \sum_{j \in \mathbb{N}} c_{i,j} g_i(u) h_j(v), \qquad c_{ij}=\langle x, g_i \otimes h_j \rangle.
        \end{equation}
        %where $c_{ij}=\langle x, g_i \otimes h_j \rangle$.
        Thanks to the existence of a bijection between $\mathbb{N}$ and $\mathbb{N}^2$ we can rearrange the terms of the basis system in order to obtain one depending on a single index instead of two, namely $\{\phi_k(u,v)\}_k$ instead of $\{g_i(u) \otimes h_j(v)\}_{i,j}$. We can thus rewrite \eqref{eqn:basis_dec_long} as:
        \begin{equation}
        \label{eqn:basis_dec}
            x(u,v) = \sum_{k \in \mathbb{N}} c_{k} \phi_k(u,v).
        \end{equation}
        In practical applications, one typically truncates the number of basis functions on the two univariate domains to $K_1$ and $K_2$ respectively, obtaining a total number of basis equal to $K:= K_1 + K_2$. 
        Let us now introduce the vector $\bm{\phi}(u,v) \in \mathbb{R}^K$, $\bm{\phi}(u,v) = [\phi_1(u,v), \dots, \phi_K(u,v)]^T$ and the
        matrix $\bm{C} \in \mathbb{R}^{T \times K}$ with elements $C[t,k]=c_{tk}$ containing the coefficients of basis projection of the random functions $Y_1, \dots, Y_T$, in such a way that:
        \begin{equation}
        \label{eqn:data_basis_dec}
            Y_t(u,v) = \sum_{k=1}^{K} c_{tk} \phi_k(u,v) + \delta_t(u,v) \qquad t=\, \dots, T,
        \end{equation}
        where $\delta_t(u,v)$ is a projection error, which is present due to the truncation of the basis system to the first $K$ terms.
        In the remainder of this section, we will neglect the projection error and identify the observed functions with the ones reconstructed from the first $M$ basis. 
        Following the work of \citealt{ramsay2005functional} and exploiting representation \eqref{eqn:basis_dec}, we aim to rephrase the eigenequation \eqref{eqn:eigenproblem_functional} in a matricial form.
        The estimated covariance function can be expressed in matrix terms:
        \begin{align*}
          \hat{\gamma}_0 (u,v;w,z) &= \frac{1}{T} \sum_{t=1}^T Y_t(u,v) Y_t(w,z) 
          = \frac{1}{T} \sum_{t=1}^T \sum_{l,m=1}^{K} c_{il} \phi_l(u,v) c_{im} \phi_m(w,z) 
          = \frac{1}{T} \bm{\phi}(u,v)^T \bm{C}^T \bm{C} \bm{\phi}(w,z). 
        \end{align*}
        Suppose now that an eigenfunction $\xi$ 
        admits the decomposition: 
        \begin{align}
            \xi(u,v) &= \sum_{l=1}^K b_l \phi_l(u,v) + \kappa(u,v) = \bm{\phi}(u,v)^T \bm{b} + \kappa(u,v),
            \label{eqn:eigen_dec}
        \end{align}
        where $\bm{b} = [b_1, \dots, b_K]^T = [\langle \xi, \phi_1 \rangle, \dots, \langle \xi, \phi_K \rangle ]^T$ contains coefficients of basis projection of $\xi$. \\
        Neglecting once again the projection error $\kappa$, the goal becomes now to estimate the coefficients $\bm{b}$ and the corresponding eigenvalue $\lambda$ for each eigenfunction $\xi_j$, for $j=\,\dots,M$.
        Let's introduce finally $\bm{W} \in \mathbb{R}^{K \times K}$, defined as $\bm{W} := \int_c^d \int_e^f \bm{\phi}(u,v) \bm{\phi}(u,v)^T du dv$, notice that the tensor product basis is composed by two orthonormal basis systems, the resulting tensor product basis system is itself orthonormal 
        % qua volendo potrei mettere la dim a piè di pagina
        and thus $\bm{W} = \bm{I}$, where $\bm{I}$ denotes the diagonal matrix. 
        The lhs of \eqref{eqn:eigenproblem_functional} can be rewritten as:
        \begin{align}
          \hat{\Gamma}_0 \hat{\xi} (u,v) &= \int_c^d \int_e^f \frac{1}{T} \bm{\phi}(u,v)^T \bm{C}^T \bm{C} \bm{\phi}(w,z) \bm{\phi}(w,z)^T \hat{\bm{b}} dw dz = \\
          &= \frac{1}{T} \bm{\phi}(u,v)^T \bm{C}^T \bm{C} \left( \int_c^d \int_e^f \bm{\phi}(w,z) \bm{\phi}(w,z)^T dw dz \right) \hat{\bm{b}} = \\
          &= \frac{1}{T} \bm{\phi}(u,v)^T \bm{C}^T \bm{C} \bm{W} \hat{\bm{b}}. %= \\
          %&= \frac{1}{T} \bm{\phi}(u,v)^T \bm{C}^T \bm{C} \hat{\bm{b}}
        \end{align}
        The eigenequation thus becomes:
        \begin{gather}
          \frac{1}{T} \bm{\phi}(u,v)^T \bm{C}^T \bm{C} \bm{W} \hat{\bm{b}} = \lambda \bm{\phi}(u,v)^T \hat{\bm{b}} \qquad \forall u,v, 
        \end{gather}
        which implies:
        \begin{gather}
          \frac{1}{T} \bm{C}^T \bm{C} \bm{W} \hat{\bm{b}} = \lambda \hat{\bm{b}}.
        \end{gather}
        We can hence derive the eigenvectors $\hat{\bm{b}}_j$ of $\frac{1}{T}\bm{C}^T\bm{C}\bm{W}$ and the corresponding eigenvalues and finally reconstruct the eigenfunctions $\hat{\xi}_j$ thanks to \eqref{eqn:eigen_dec}.
        
    %%%%%%%%%%%%%%%%%%%%%%%%%%%%%%%%%%%%%%%%%%%%%%%%%%%%%%%%%%%%%%%%
    
    \subsection{Comparison of Estimation Methods}
    \label{sec:comparison_FPCA}

    In this section, we aim to compare the two proposed approach for performing functional principal component analysis, namely the basis expansion and the discretization approach. 
    Without loss of generality, we settle the study in $\mathcal{L}^2([0,1]\times[0,1])$.
    
    In general, given a finite basis system $\{\phi_k\}_{k=1,\dots,K}$, one can represent a Functional Time Series $\{Y_t\}_{t=1}^{T}$ by means of representation \eqref{eqn:data_basis_dec}, that we report here for ease of reference:
    \begin{equation}
    \label{eqn:basis_decomposition_2}
        Y_t(u,v) = \sum_{k=1}^{K} c_{tk} \phi_k(u,v) + \delta_t(u,v) \qquad t=1, \dots, T.
    \end{equation}
    Starting from this decomposition, we have derived in \ref{sec:FPCA_basis} an estimator of the functional principal components $\xi_j$, which, however, neglects the contribution of the error $\delta_t$. For such reason, the choice of the basis system $\{\phi_k\}_{k=1}^K$ is crucial, and if a meaningful option is not available, the approximation residual $\delta_t$ would be large and this procedure would inevitably provide biased estimates of $\xi_j$.
    On the other hand, the FPCA approach based on data discretization provides good result as long as the sampling grid is sufficiently dense. 
    
    In order to prove such thesis, we simulate a time series of functions $\{Y_t\}_{t=1}^T \subset$ span$\{\phi_1, \dots, \phi_K\}$, from a non-concurrent FAR(1) process with Gaussian errors. Data are simulated based on a basis expansion on $\{\phi_1, \dots, \phi_K\}$, which is constructed as the tensor product basis of two Fourier basis systems $\{g_i\}_{i=1,\dots,K_1}$, $\{h_i\}_{i=1,\dots,K_2}$ both defined on $[0,1]$. The sample size is chosen equal to $T=50$, the number of basis in each of the one-dimensional systems is selected equal to 5, in order to have a total number of basis $K=25$.
    
\begin{figure} 
        \centering
        \begin{subfigure}[b]{0.75\textwidth}
            \centering \includegraphics[width=1\linewidth]{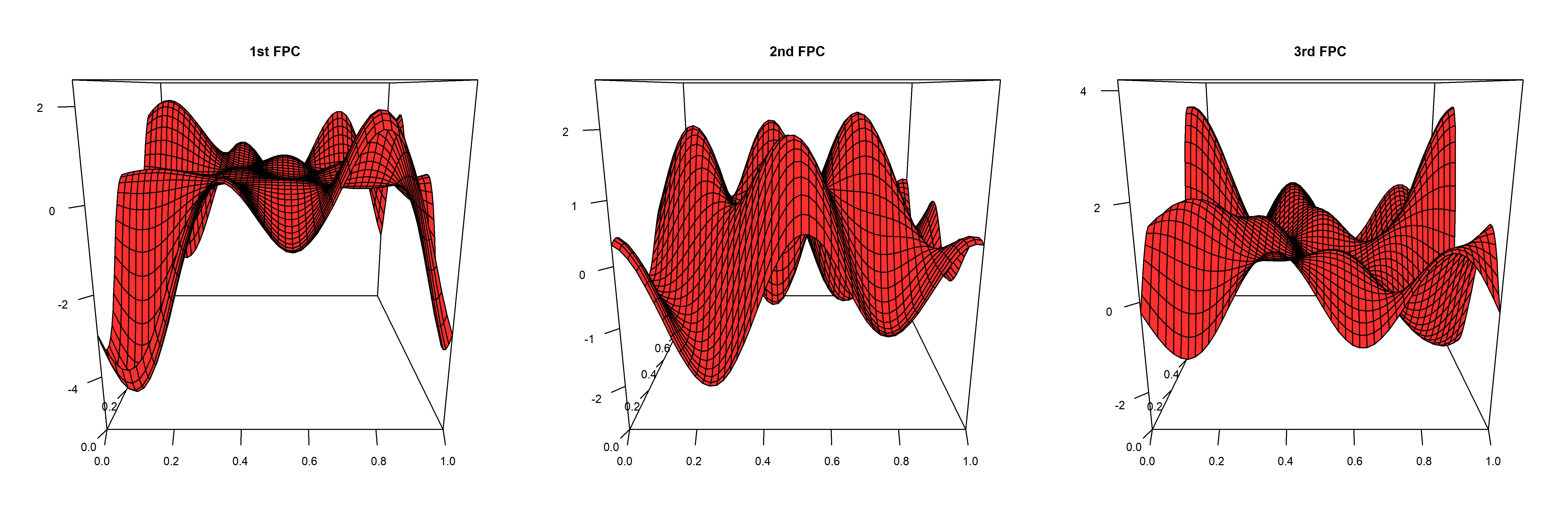}
            \caption{FPCA by basis expansion on the same basis system used for the simulation.}
            \label{fig:FPCA_comparison_1}
        \end{subfigure}
        \begin{subfigure}[b]{0.75\textwidth}
            \centering \includegraphics[width=1\linewidth]{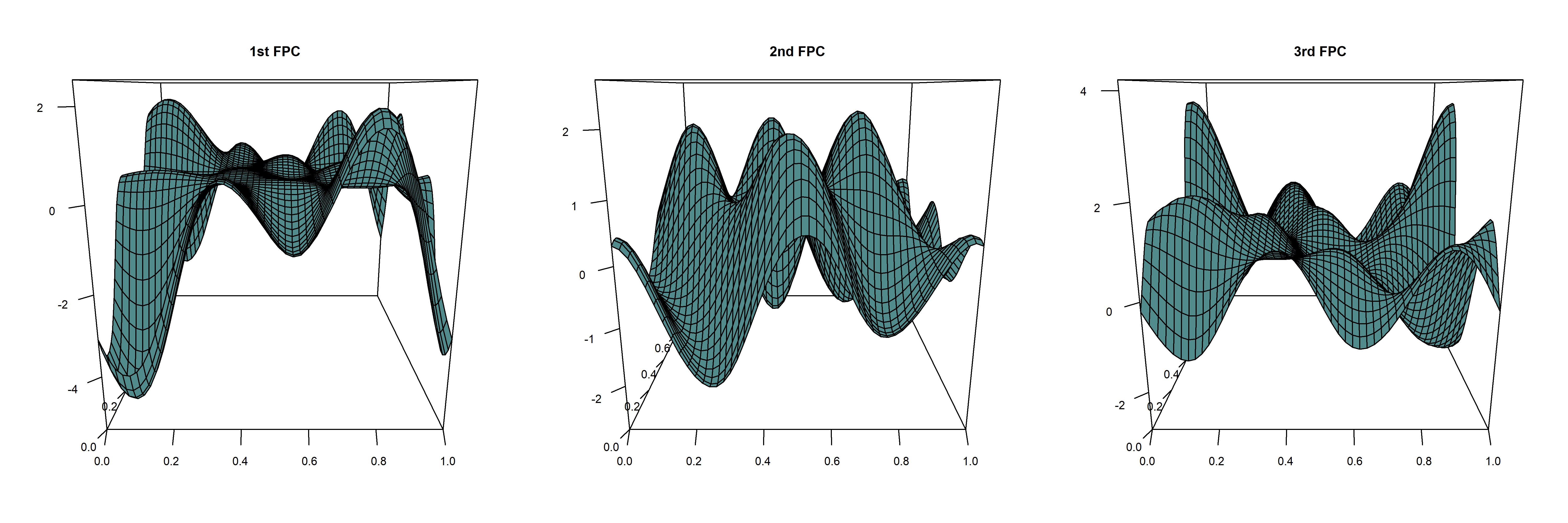}
            \caption{FPCA by basis expansion on a different basis system.}
            \label{fig:FPCA_comparison_2}
        \end{subfigure}
        \begin{subfigure}[b]{0.75\textwidth}
            \centering \includegraphics[width=1\linewidth]{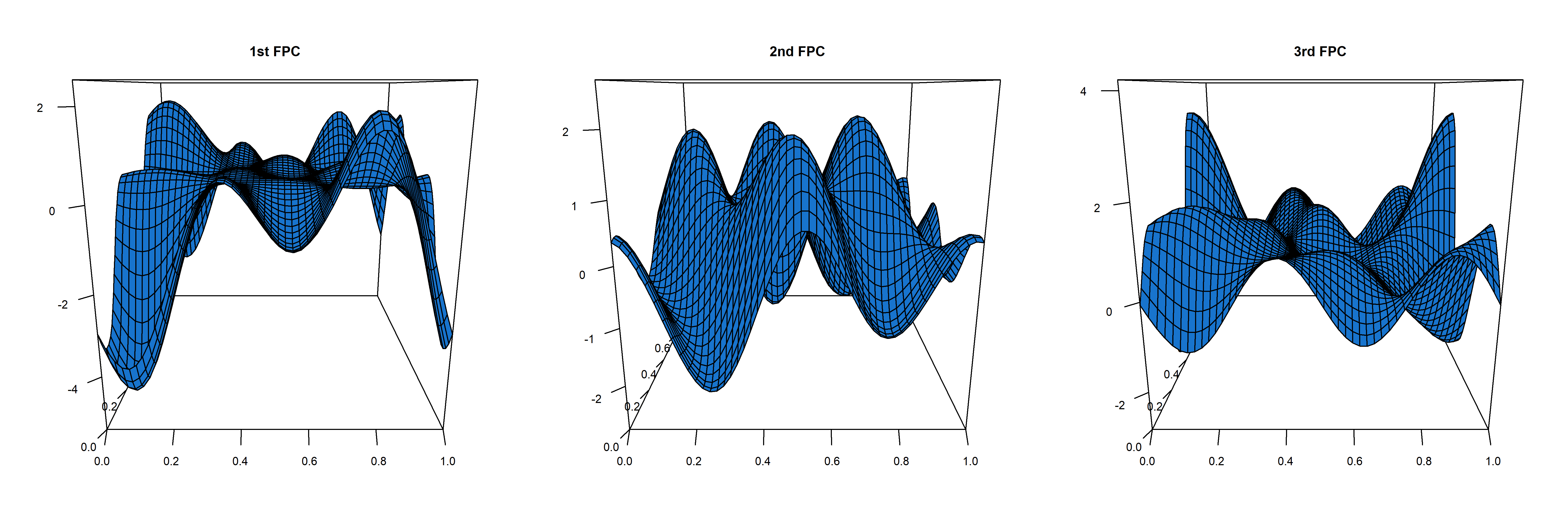}
            \caption{FPCA by grid discretization.}
            \label{fig:FPCA_comparison_3}
        \end{subfigure}
        %\caption{First three functional principal components estimated with three different methods. As usual in PCA, EFPC's are unique up to a constant. For such reason, in order to compare the different approaches, estimated harmonics in the second and third row are each rescaled by means of the mean difference ratio with the ones in the first row.}
        \caption[First three functional principal components, estimated with different methods]{First three functional principal components, estimated with three different methods. As usual in PCA, EFPC's are unique up to a constant. For such reason, in order to compare the different approaches, estimated harmonics in the second and third row are each rescaled by means of the mean difference ratio with the ones in the first row.}
        \label{fig:}
    \end{figure}

    Since we know the space where the functions are embedded, we can apply the estimation procedure in \ref{sec:FPCA_basis} using as basis system the same one used in the simulation. Notice that in this case the approximation error $\delta_t$ in \eqref{eqn:basis_decomposition_2} is exactly zero and one can derive optimal estimates of the functional principal components.
    Estimators of the first three functional principal components are represented in \autoref{fig:FPCA_comparison_1}. 
    We repeat the same estimation procedure, this time modelling functions with a basis built from the tensor product of two one-dimensional cubic B-Spline basis systems with 5 basis each. In this case, the basis system do not coincide with the one from which functions are simulated. Nevertheless, as reported in \autoref{fig:FPCA_comparison_2} all the scaled eigenfunctions are very close to the optimal ones estimated before. %The mean squared approximation error is reported in ??, and, as one could expect, it tends to decrease as the number of basis functions increases.
    %Despite having chosen a large number of basis functions, in this case we are not able to reconstruct eigenfunctions other then the first one. %Whereas the first eigenfunction is estimated quite well by this second approach, the quality of the other estimates are very poor as reported in \autoref{fig:FPCA_comparison_2}
    %As reported in \autoref{fig:FPCA_comparison_2}
    %whereas the first eigenfunction is estimated quite well in this second scenario, the quality of the other estimates are very poor despite the high number of basis used to model the process.
    Finally, we compare the aforementioned estimators with the ones coming from the discretization approach. Each of the one-dimensional grids is discretized using a step size equal to 0.02, thus resulting in a total of 2500 points. Also in this case (see \autoref{fig:FPCA_comparison_3}), estimated harmonics are very close to the optimal ones in \autoref{fig:FPCA_comparison_1}.
    
    To enable for better comparison, we report in \autoref{table:MSE_FPCA} the Mean Squared Error (MSE) \eqref{eqn:MSE} between the FPC's $y$ estimated using full knowledge of the basis system from which functions are simulated and the ones estimated using other techniques ($\hat{y}$). Such quantity is computed starting from values of $y$ and $\hat{y}$ on a two-dimensional grid $\{(u_i,v_j)\}_{i=1,\dots,N_1;j=1,\dots,N_2}$.
    \begin{equation}
    \label{eqn:MSE}
        MSE(y,\hat{y}) = \frac{1}{N_1 N_2} \sum_{i=1}^{N_1} \sum_{j=1}^{N_2} \left(y(u_i,v_j) - \hat{y}(u_i,v_j)\right)^2.
    \end{equation}
    We can appreciate very low values of MSE, regardless of the technique used for FPCA, thus suggesting that both the basis expansion and the discretization approach are valid options on a practical point of view.
    
    \begin{table}
        \centering
        \begin{tabular}[t]{lccc}
                \toprule
                FPCA method &  \multicolumn{3}{c}{MSE} \\
                \midrule
                & 1st FPC      & 2nd FPC     & 3rd FPC    \\
                \midrule
                Basis on B-Spline & $2.62 \cdot10^{-3}$ & $4.45\cdot 10^{-3}$ & $2.28 \cdot10^{-3}$ \\
                Discretization    & $7.69 \cdot 10^{-4}$ & $2.45\cdot 10^{-3}$ & $1.38\cdot 10^{-3}$ \\
                \bottomrule
                \end{tabular}
                \caption[Comparison of estimated FPC's]{MSE between estimated FPC's on the basis system used for the simulation and FPC's estimated using another basis system or the discretization approach.}
        \label{table:MSE_FPCA}
    \end{table}

\end{document}